\documentclass[a4paper,12pt]{article}
\RequirePackage[OT1]{fontenc}
\RequirePackage[fleqn]{amsmath}
\usepackage{graphicx}
\usepackage{amssymb}
\usepackage{booktabs}
\usepackage{xcolor,colortbl}
\usepackage[latin1]{inputenc}
\usepackage{makeidx}         
\usepackage{graphicx}        
\usepackage{multicol}        
\usepackage[bottom]{footmisc}
\usepackage{natbib}
\usepackage{caption} 
\usepackage{multirow} 
\usepackage[normalem]{ulem} 

\newtheorem{Theorem}{Theorem}

\newtheorem{Corollary}{Corollary}
\newcommand{\ind}{\mbox{$\perp\!\!\!\perp \,$}}
\DeclareMathOperator*{\Diag}{Diag}

\begin{document}

\begin{center}
{\Large\ \textbf{Hierarchical Marginal Models\\ with Latent Uncertainty} } \\ \ \\ 
$\text{Roberto Colombi}^1$,\ $\text{Sabrina Giordano}^2$,\ $\text{Anna Gottard}^3$, \ $\text{Maria Iannario}^4$\\  \ \\
\small{
\textit{$^1$ Department of Management, Information and Production Engineering, University of Bergamo, Italy,  e-mail: colombi@unibg.it}\\
\textit{$^2$ Department of Economics, Statistics and Finance, University of Calabria, Italy, e-mail: sabrina.giordano@unical.it}\\
\textit{$^3$ Department of Statistics, Computer Science, Applications, University of Florence, Italy,  e-mail:gottard@disia.unifi.it }\\
\textit{$^4$ Department of Political Sciences, University of Naples Federico II, Italy, e-mail:maria.iannario@unina.it }
}
\end{center}



\newcommand{\tr}{^{\prime}}

\def\b#1{\mbox{\boldmath $#1$}}    
\def\cg#1{\ensuremath{\mathcal{#1}}}      
\def\cgl#1{\mbox{\scriptsize {${\cal #1}$}}}

\renewcommand{\baselinestretch}{1.2}

\abstract{
In responding to rating questions, an individual may
give answers either according to his/her knowledge/awareness or to his/her level of
indecision/uncertainty, typically driven by a response style. As ignoring this dual behaviour
may lead to misleading results, we define a multivariate model for ordinal rating responses,  
by introducing, for every item, a binary latent variable that discriminates aware from uncertain responses. 
 Some independence assumptions among latent and observable variables characterize the uncertain behaviour and make the model easier to interpret. Uncertain responses are  modelled by specifying probability distributions that can depict different response styles characterizing the uncertain raters. A marginal parametrization allows a simple and direct interpretation of the parameters in terms of association among aware responses and their dependence on explanatory factors.
The effectiveness of the proposed model is attested through an application to real
data and supported by a Monte Carlo study.} 

\vspace{0.6 cm}




\section{Introduction}

When people are invited to express their opinion about a set of items by choosing among ordinal categories,  their answers  can be  either the  exact expression of their opinion or can correspond to a response style ensued from indecision or uncertainty. The first type of answers is of interest when one is focused on the \emph{true} perceived value of the items, the second type is investigated mainly in sociological and psychological studies on the uncertainty in the process of responding. 
Hence, we call \emph{awareness},  the exact expression of personal opinion on an item, and  \emph{uncertainty}, the difficulty of choosing among the ordered alternatives due to response styles, careless, unconsciousness, indecision or randomness. 
Uncertain respondents may have particular response styles, using only a few of the given options: someone can have a tendency to select the end points, others the middle categories, or only the positive/negative side of the rating scale.  Examples  are described, among many, in \cite{yates1997}, \cite{baumgartner2001},  and \cite{luchini2013}. Such response styles can potentially distort the reliability and validity of the data analysis.
%
%

To take into account the two possible behaviours in answering,  for every observable  variable  $R_i$,  rating the item $i$,  $i=1,2,\dots,v$,  we introduce a binary latent variable $U_i$ such that the conditional distribution of $R_i$ given $U_i=0$ models uncertain responses while, given $U_i=1$, it describes aware responses. The  latent variables define $2^v$ latent classes, each one corresponding to a subset $S$ of the responses such that
\begin{itemize}
	\item[i)] the observable variables belonging to $S$  are uncertain responses and the remaining ones are aware responses,
	\item[ii)] the observable variables in $S$  are mutually independent and independent of the variables not in $S$,
	\item[iii)] the distribution of the variables  in $S$ is a marginal of the distribution in the latent class  without any aware responses,
	\item[iv)] the distribution of the variables not in $S$ is a marginal of the distribution in the latent class  with only aware responses,
	\item[v)] uncertain responses are modelled through probability functions that can depict different response styles in the process of answering.
\end{itemize}

As a consequence of points iii) and iv), the marginal distributions of uncertain and aware  responses, respectively, are replicated in different latent classes. For this reason, it is convenient to  parameterize the joint distribution of the observable and latent variables through  marginal models  \citep{bergsma2002, bartolucci2007}. The marginal parametrization facilitates the interpretation of the results,  defining directly the marginal distributions of the responses in case of awareness and uncertainty and their association structure. Moreover, this parametrization greatly simplifies the inclusion of explanatory variables and  the maximum likelihood estimation.
	
We call this model \emph{Hierarchical Marginal Model with Latent Uncertainty components} (HMMLU). It  permits, for each item,  to determine the probability of an  uncertain response and to describe its dependence on individual characteristics and on the uncertainty in other items. An HMMLU enables to distinguish the distribution of responses dictated by the awareness from those dictated by a response style due to uncertainty.

A variety of non-model-based and model-based procedures have been provided to detect and control for the effect of response styles in rating data. Non-model-based approaches \cite[e.g.][]{meade2012} include techniques aimed at detecting uncertain responses. For instance, they use indicators such as frequency accounts of endpoint responses or the computation of the standard deviation of item scores within a respondent. According to these methods, inattentive respondents are identified and usually excluded from the analysis. Therefore, this kind of procedures essentially ends up with a data cleaning process, whose results may be strongly influenced by the adopted indices for screening for unreliable responses.
	
Model-based procedures that  present  similarities with our proposal are item response theory (IRT) models for ordinal responses  and latent class factor (LCF) models, which involve a latent variable  that directly affects all the observable variables to account for uncertainty. 
In particular, such models assume that a multidimensional latent trait underlies item responses,  and that the items are locally independent when the latent trait levels and the response style are controlled for. 
In this context, \cite{jin2014}, \cite{huang2016} and \cite{tutz2018},  among others, propose random threshold IRT models for polytomous variables where the response style is included,  in different ways,  as a random effect. This random component reduces or increases the distance between thresholds so that the extreme (middle) categories are more likely to be endorsed. \cite{bockenholt2017} \citep[see also][]{von2007mixture} present an IRT model that allows for heterogeneity of thresholds across latent classes. Other authors \citep[e.g.][]{morren2011}  provide LCF models where the response style is a discrete ordinal latent variable. 

The model we are proposing presents some advantages on the aforementioned approaches, that can be sketched in a few key points.
Firstly, we assume that every observable variable is driven by its own binary latent variable to account for a item-specific uncertainty. Consequently, we identify subgroups of respondents who can exhibit different uncertainty/response styles for subsets of items. This is not possible when only one latent variable affects all the items. 	
Defining $2^v$ latent with a univocal meaning permits to distinguish uncertain and aware responses for every subset of items. The other approaches based on mixture models need to select the number of latent classes and make a subjective interpretation of their meaning. 
In addition, our proposal replaces the local independence hypothesis with 
condition ii). That is to say, 
only the uncertain responses are assumed independent, while association among aware responses is modelled directly. 
		
In addition, uncertain responses are in our paper explicitly modelled by probability functions with flexible (Uniform, bell and U) shape that can take into account both randomness and tendency to the extreme or middle categories. Subgroups of uncertain respondents can also have different response styles. In other approaches, these distributions are not directly modelled.
Finally, a marginal modelling in our proposal allows a simpler and direct interpretation of the parameters in terms of marginal distributions of aware responses and in terms of their association. Other approaches deal with association only indirectly, assuming independence given the latent variables.  

In our opinion, the here proposed model enriches the literature with new perspectives and useful advantages.  
	
The rest of the paper is organized as follows. We firstly present  the model in the bivariate case (Sec.~\ref{Sec.bivariate}) to exemplify our proposal in a simple setting. We discuss the general case in Sections \ref{Sec.hierarch}, describing the main assumptions (Sec.~\ref{Sec.assumptions}, \ref{Sec.assumptions2}), the parameterization adopted and identifiability issues  (Sec.~\ref{Sec.parameterization}, \ref{Sec.ident}). The bias in the parameter estimates introduced  by ignoring uncertainty in the answers is illustrated  in Section \ref{MonteCarlo} via a Monte Carlo study.
An application  and some concluding remarks are provided in Sections \ref{example} and \ref{Conclusion}, respectively. Analytical technicalities  are reported in the two  Appendices.

\section{A mixture model for two responses}\label{Sec.bivariate}
For clarity, it is useful to introduce the main features of our model in the simple case of two items and delay the general presentation to the next section.

Let $R_1$ and $R_2$ be two ordinal variables with support $\{1,2,\dots,m_1\}$ and $\{1,2,$ $\dots,m_2\}$, respectively.
We assume the existence of two binary latent variables,  $U_l$, $l=1,2$, such that the respondent  answers the $l^{th}$ question according to his/her
awareness when $U_l=1$ or his/her uncertainty when $U_l=0$.

The  joint distribution of the observable variables is specified by the mixture
\begin{equation} \label{Imistura}
P(R_1 = r_1, R_2 = r_2) = \sum_{ij \in \{0,1\}^2} \pi_{ij} \, P(R_1 = r_1, R_2 = r_2 \mid U_1 = i, U_2 = j)  \\ \end{equation}
for every $r_1 = 1,2,\dots,m_1$ and $r_2 = 1,2,\dots,m_2,$
where  $\pi_{ij} = P(U_1 = i, U_2 = j)$, $i=0,1$, $j=0,1$, are the joint probabilities of the latent variables. Specifically  they are the probabilities that both the answers are given with awareness ($\pi_{11}$), both with uncertainty ($\pi_{00}$) or one with uncertainty and the other one with awareness ($\pi_{01}$ and $\pi_{10}$).

To adapt this general mixture to the particular task of allowing for individual uncertainty in responding, we introduce some further assumptions, also  important to substantially simplifying model \eqref{Imistura}. These assumptions are consistent with the idea that uncertain responses are driven by randomness.

We assume that each observable variable $R_l$  depends only on its latent variable $U_l$, $l=1,2,$ i.e.
\begin{equation} \label{A}
R_1 \ind U_2 \mid U_1, \quad R_2 \ind U_1 \mid U_2, \end{equation}
and that the observable responses $R_1$ and $R_2$ are independent when at least one of them is given under uncertainty. Therefore,
\begin{equation}\label{AA}
R_1 \ind R_2 \mid U_1=0, U_2=0, \:R_1 \ind R_2 \mid U_1=0, U_2=1, \: R_1 \ind R_2 \mid U_1=1, U_2=0.
\end{equation}
Consequently, mixture \eqref{Imistura} simplifies to
\begin{equation} \label{mistura}
\begin{array}{rcl}
P(R_1 = r_1, R_2 = r_2)  & = & \pi_{00} \, g_1(r_1) \, g_2(r_2) \\
&  & + \,\pi_{01} \, g_1(r_1) \, P(R_2=r_2\mid U_2 = 1) \\
  &  & + \, \pi_{10} \,P(R_1=r_1\mid U_1 = 1) \, g_2(r_2) \\ &  & +
\pi_{11} \, P(R_1 = r_1, R_2 = r_2 \mid  U_1 = 1, U_2=1).
\end{array}
\end{equation}
In \eqref{mistura}, $P(R_l=r_l\mid U_l = 1)$ is the distribution of the aware responses, with $r_l=1,2,\dots,m_l$  and $g_l(r_l)=$ $P(R_l=r_l\mid U_l = 0)$, $l=1,2$, denotes the distribution of responses under uncertainty.

An important consequence of assumption \eqref{A} on the specification of model \eqref{mistura} is that it imposes coherence in the marginal distributions.

In fact, it ensures that marginalizing the 
distribution $P(R_1 = r_1, R_2 = r_2 \mid  U_1 = 1, U_2=1)$  of the two responses in the last component of \eqref{mistura} over $R_1$ (or $R_2$),  one get exactly the   
distributions of the aware responses, $P(R_1=r_1\mid U_1 = 1) = P(R_1 = r_1\mid U_1=1, U_2=0)$  (or $P(R_2=r_2\mid U_2 = 1) =P(R_2 = r_2\mid U_1=0, U_2=1)$), involved in the second (third) component of equation \eqref{mistura}.


To facilitate the interpretation and the maximum likelihood estimation of the parameters, it is convenient  to introduce  a marginal parameterization \citep{bergsma2002} for the mixture \eqref {mistura}.
The distribution of the latent variables $(U_1, U_2)$, defined in \eqref {mistura} by the probabilities $\pi_{ij}$, $i=0,1$, $j=0,1$, is parameterized through a marginal logit for each latent variable, measuring the probability of being uncertain on each specific item, plus a log odds ratio. 
When this parameter is positive,  respondents tend to have the same behaviour of uncertainty/awareness on the two items.

To parameterize the $m_1$ probabilities $P(R_1 = r_1\mid U_1=1)$, and the $m_2$ probabilities $P(R_2 = r_2\mid U_2=1)$, we define $(m_1 - 1)$ and  $(m_2 - 1)$ logits, chosen from local, global, continuation, or reverse continuation logits.
These logits, together with $(m_1 - 1)(m_2 - 1)$ log odds ratios (local, global, continuation, reverse continuation), are used to parameterize the joint distribution $P(R_1 = r_1, R_2 = r_2 \mid U_1=1, U_2=1)$. 

Unfortunately, even when the uncertainty distributions $g_l(r_l)$,
$r_l=1,2,\dots,m_l,$ do not depend on unknown parameters, the model includes  $m_1m_2 - 1 + 3$  parameters. Therefore, identifiability constraints are necessary.
For instance, under the constraint of uniform association that imposes identical log odds ratios, the number of parameters $m_1+m_2+2$ does not exceed $m_1m_2-1$, the number  of independent observable frequencies,  when $m_1,m_2 \ge 3$.
In addition, the presence of covariates  may also help to make  the model identifiable, as will be shown in Section \ref{Sec.ident}. 

Regarding the uncertainty distributions, the simplest choice not depending on unknown parameters  is the  discrete Uniform distribution,
previously used by \cite{delia2005} 
to  model  uncertainty in the univariate case. Several more realistic distributions, not depending on any parameter, have been discussed by \cite{gottard2016}.   A more flexible distribution with a shape parameter for describing  different response styles will be proposed in Section~\ref{Sec.parameterization}.

\section{A mixture model for more than two responses}\label{Sec.hierarch}
In this section, we introduce the class of Hierarchical Marginal Models with Latent Uncertainty (HMMLU) that generalizes the model of Section \ref{Sec.bivariate}  to the case of more than two responses.

Given $v$ ordinal variables $R_i$, with categories $r_i = 1,2,\dots,m_i$, $i=1,2,\dots,v$, the vector  $\b r=(r_1,r_2,\dots,r_v)$ will denote one of their $m = \prod_{i=1}^v m_i$ possible joint realizations.
To model uncertainty in answering, we assume the existence of  $v$ latent dichotomous variables $U_i$, $i=1,2,\dots,v$,  whose joint realizations $\b u=(u_1,u_2,\dots,u_v)$ are called \emph{uncertainty configurations}. In an uncertainty configuration, a $0$ in the $i^{th}$ position stands for an uncertain behaviour in answering the $i^{th}$ question.
Hereafter, $p(\b r|\b u)$ will denote the distribution of the  observable variables given the latent ones and $\pi(\b u)$ the joint distribution of the latent variables.
Consequently, the joint distribution of the observable variables is the  mixture
\begin{equation} \label{mixt}
p(\b r)=\sum_{\b u \in  \{0,1\}^v}p(\b r \: | \: \b u) \pi(\b u), \end{equation}
of $2^v$ components corresponding to the uncertainty configurations $\b u=(u_1,u_2,\dots,u_v)$, analogous to \eqref{Imistura} given in the bivariate case.

This model is well specified only by adding some assumptions, that will be introduced in Section \ref{Sec.assumptions}.
To this aim, further notation is required.
Given the set of indices $\cgl V=\{1,2,\dots,v\}$, let $\cg R=\{R_i: i \in \cgl V\}$ and $\cg U=\{U_i: i \in \cgl V \}$  denote the set of observable  and  latent variables,  respectively.
For every $\cgl S \subset \cgl V$, we  specify the subsets $\cg R_{\cgl S}=\{R_i: i \in \cgl S\}$ and $\cg U_{\cgl S}=\{U_i: i \in \cgl S \}$. Specifically, for every uncertainty configuration $\b u$, we will be interested in the subset of indices $\cgl  V(\b u)=\{i : u_i=0, i \in \cgl V\}$  and the subset $\cg R_{\cgl V(\b u)}$ of variables observable under uncertainty. Moreover, it will be useful  the configuration $\b u^*=\b 1_{v}$ of no uncertain responses, i.e.\  $\cgl V(\b u^*)=\emptyset.$
Finally, for each $\b r$, $\b u$, we will denote  with $\b r_{\cgl S}$ and $\b u_{\cgl S}$ the marginal  configurations of the variables in $\cg R_{\cgl S}$ and $\cg U_{\cgl S},$ respectively.
For the sake of  simplicity,  we will use the shorthand notation $p(\b r_{\cgl S})$ to indicate the marginal probabilities  $p_{_{\cgl R_{\cgl S}}}(\b r_{\cgl S})$ and $p(\b r_{\cgl S}\mid \b u_{\cgl T})$ to indicate the conditional probabilities $p_{_{ \cgl R_{\cgl S} \mid \; \cgl U_{\cgl T}}}(\b r_{\cgl S} \mid \b u_{\cgl T}).$

The proposed model will contemplate  heterogeneity if $p(\b r \mid \b u)$ and $\pi(\b u)$ vary according to subject's characteristics. We will clarify how to model the effect of covariates in Section~\ref{Sec.parameterization}, where respondents are grouped in strata identified by some  covariate patterns. For simplicity, we will consider discrete explanatory variables only. Continuous covariates may be also taken into account.

\subsection{Model assumptions}\label{Sec.assumptions}

To characterize the awareness/uncertainty attitude in giving answers,  we make the following assumptions that generalize those given in Section \ref{Sec.bivariate}  to the case of $v$, $v>2$, responses. These assumptions formalize  the idea that uncertainty implies randomness in responding,   that  a specific latent variable    is needed for every item to account for uncertainty  and that, for every respondent, an  uncertain answer  is independent of all the other (uncertain or aware) responses.

\begin{description}
\item[Assumption A1:] \emph{Specific latent variables}

For every $\cgl S \subset \cgl V$,
$$ \cg R_{\cgl S} \ind \cg U_{ \:\cgl V \: \setminus \cgl S}  \mid \cg U_{\cgl S}.$$
\end{description}
With respect to  Section \ref{Sec.bivariate},  A1 generalizes \eqref{A} and implies that every subset $\cgl S$ of observed variables depends on its corresponding subset of latent variables. Equivalently,
$p(\b r_{\cgl S}|\b u)= p(\b r_{\cgl S}|\b u_{\cgl S})$ for every $\b r$, $\b u$ and $\cgl S \subset \cgl V$.
\begin{description}
\item[Assumption A2:] \emph{Context specific independence due to uncertainty}

For every  configuration $\b u$ and every $\cgl S \subseteq \cgl V(\b u)$,
\begin{align*}
\cg R_{\cgl S} &\ind \cg R_{\cgl V(\b u) \: \setminus \: \cgl S}  \mid \b u,\\
\cg R_{\cgl V(\b u)} &\ind  \cg R_{\cgl V \: \setminus \: \cgl V(\b u)} \mid \b u.
\end{align*}

\end{description}
These independences are context specific \citep{H2004} as they hold given a specific configuration $\b u$, with the set of variables involved changing with $\b u$.
Assumption  A2 generalizes  \eqref{AA}.
In particular, the first statement implies that, conditionally  on  $\b u$, the variables in $\cg R_{\cgl V(\b u)}$, describing uncertain responses,  are mutually independent. The second statement says that, conditionally  on  $\b u$, the variables in $\cg R_{\cgl V(\b u)}$ are independent of the remaining observable variables.

The next assumption is needed to facilitate the identifiability  of the parameters of the model and their interpretation.

\begin{description}
	\item[Assumption A3:] \emph{Composition property}
	
For every $\cgl S \subset \cgl V$, $\cgl T \subset \cgl V$, $\cgl S \cap \cgl T = \emptyset$,
\begin{center}
		$\cg R_{\cgl S} \ind \cg R_{\cgl T} \mid \b u^*$ is equivalent to $\cg R_{i} \ind \cg R_{j} \mid \b u^*$ for every $i \in \cgl S$ and $ j \in \cgl T$.
\end{center}

\end{description}
This assumption  states that the probability function $p(\b r \: | \: \b u^*)$ has to satisfy the composition property of conditional independence \citep[][page 33]{studeny2005}.
This property is not generally valid. By \citet[Lemma 1]{lupparelli2009} or \citet[Lemma 1]{kauermann1997}, it is equivalent to requiring that $p(\b r \: | \: \b u^*)$ has all the Glonek-McCullagh interactions  \citep{glonek1995} among more than two variables equal to zero.
To understand the usefulness  of this  assumption, note that it makes sense only for $v>2$. For $v=3$ it implies that $(m_1-1)(m_2-1)(m_3-1)$ three-way interactions are null  in the joint distribution of the  responses given that $U_i=1, \: i=1,2,3.$ These restrictions, when $m_i \geq 3,\: i=1,2,3,$  allow for the introduction of the $7$ parameters needed to fully parameterize the joint distribution of the three binary latent variables.
For $v>3$ or in the presence of covariates, Assumption A3 may be too restrictive or unnecessary and can be  relaxed. However, it has the advantage of 
enhancing the interpretability of the model.

\subsection{Consequences of model assumptions} \label{Sec.assumptions2}

The following theorems highlight some important features of the proposed model that are consequences of the assumptions in Section \ref{Sec.assumptions}.
In particular, Theorem \ref{teo1} plays a key role in the model specification, showing how the components of mixture \eqref{mixt} simplify according to A1 and A2.

\begin{Theorem} \label{teo1} Assumptions  A1 and A2 imply that
\begin{equation}\label{pru} p(\b r \mid \b u) \, = \, p(\b r_{\cgl V \setminus \cgl V(\b u)}\mid \b u_{\cgl V \setminus \cgl V(\b
u)})\prod_{i \in \cgl V(\b u)}g_i(r_i), \end{equation}
 where $g_i(r_i)$ are the marginal probabilities $P(R_i=r_i\mid U_i = 0)$ of the uncertain responses, $i \in \cgl V(\b u)$.
Moreover, the  joint distributions of aware responses $p(\b r_{\cgl V \setminus \cgl V(\b u)}\mid \b u_{\cgl V
\setminus \cgl V(\b u)})$ are  marginal distributions of $p(\b r \mid \b u^*)$.
 \end{Theorem}

\emph{Proof.} Equation \eqref{pru} derives from  $p(\b r \mid\b u) = p(\b r_{\cgl V \setminus \cgl V(\b u)}\mid \b u) \: p(\b r_{ \cgl V(\b u)}\mid \b r_{\cgl V \setminus \cgl V(\b u)},\b u)$. The first factor of the last product simplifies to $p(\b r_{\cgl V \setminus \cgl V(\b u)}\mid \b u)=p(\b r_{\cgl V \setminus \cgl V(\b u)}\mid \b u_{\cgl V \setminus \cgl V(\b u)})$ due to A1. For the second factor, it is  $p(\b r_{ \cgl V(\b u)} \mid \b r_{\cgl V\setminus \cgl V(\b u)},\b u)=p(\b r_{ \cgl V(\b u)}\mid \b u_{ \cgl V(\b u)})$ according to A1 and the second statement of A2. The first part of the thesis follows since $p(\b r_{ \cgl V(\b u)}\mid \b u_{ \cgl V(\b u)})$ factorizes in the product of
marginal probabilities $g_i(r_i)$ as a consequence of the first statement of Assumption A2.
The second part of the thesis is proved by noting that the equality $p(\b r_{\cgl V \setminus \cgl V(\b u)}\mid \b u^*) = p(\b r_{\cgl V \setminus \cgl V(\b u)}\mid \b u^{*}_{\cgl V
\setminus \cgl V(\b u)})$  is true according to Assumption A1
and because of $\b u^{*}_{\cgl V
\setminus \cgl V(\b u)}=\b u_{\cgl V
\setminus \cgl V(\b u)}.$ \hfill $\Box$

\vspace{0.5cm}

The following corollaries of Theorem \ref{teo1} clarify the independence structure among observable variables implied by the proposed model.
%

In particular, the  corollaries entail that  independences among the observable variables, holding conditionally on the configuration of no uncertainty $\b u^*$, are also valid conditionally on other configurations (Corollary \ref{cor1}) and, with further assumptions, unconditionally (Corollary \ref{cor2}).

\begin{Corollary}\label{cor1}
	Suppose A1 and A2 hold.
\begin{itemize}
\item[$i)$] If $\;\cg R_{\cgl S} \, \ind \,  \cg R_{\cgl T} \mid \b u^*$, with  $\cgl S \subset \cgl V$, $\cgl T \subset \cgl V$ and  $\cgl S \cap \cgl T = \emptyset$, then $\cg R_{\cgl S} \: \ind \: \cg R_{\cgl T}\mid \cg U$.
\item[$ii)$]  If  $\; \cg R_{\cgl S} \ind \cg R_{\cgl T} \mid \cg (R_{\cgl W}, \, \b u^*)$, then  $\cg R_{\cgl S} \ind \cg R_{\cgl T} \mid (\cg R_{\cgl W}, \,\b u)$
for every $\b u$ such that $\cgl W \subseteq \cgl V \setminus \cgl V(\b u)$.
\end{itemize}
\end{Corollary}
Corollary \ref{cor1} directly derives  from Theorem \ref{teo1} by applying the factorization criterion.
%
%

\begin{Corollary} \label{cor2}
Suppose A1 and A2 hold. If $\cg R_{\cgl S} \, \ind \,  \cg R_{\cgl T} \mid \b u^*$ and  $\cg U_{\cgl S} \, \ind \, \cg U_{\cgl T}$, with  $\cgl S \subset \cgl V$, $\cgl T \subset \cgl V$ and  $\cgl S \cap \cgl T = \emptyset$, then $\cg R_{\cgl S} \: \ind \: \cg R_{\cgl T}$.
\end{Corollary}

\emph{Proof.}
According to Corollary \ref{cor1}, it holds that $\cg R_{\cgl S} \ind \cg R_{\cgl T} \mid \cg U$. By Assumption A1,  it results
$\cg R_{\cgl S} \: \ind \: \cg U_{\cgl V \setminus \cgl S \: \cup \: \cgl T} \mid  \cg U_{\cgl S \: \cup \: \cgl T}.$
By the contraction  property of conditional independence \citep{studeny2005}, the previous two independences are equivalent to $\cg R_{\cgl S} \ind \cg R_{\cgl T}, \: \cg U_{\cg V \setminus \cgl S \: \cup \: \cgl T}\mid  \cg U_{\cgl S \: \cup \:  \cgl T}$, which implies
\begin{equation} \cg R_{\cgl S} \ind \cg R_{\cgl T}\mid \cg U_{\cgl S \: \cup \: \cgl T}. \label{four}\end{equation}
Moreover,  A1 ensures the following independences
\begin{equation}\label{three}
i:\quad
\cg R_{\cgl S} \ind \cg U_{\cgl T}\mid \cg U_{\cgl S}, \qquad \qquad
ii:\quad
\cg R_{\cgl T} \ind \cg U_{\cgl S} \mid \cg U_{\cgl T}.
\end{equation}
Now, applying the contraction property to \eqref{four} and (\ref{three}$i$), we obtain
\begin{equation}
 \cg R_{\cgl S} \ind (\cg R_{\cgl T}, \cg U_{\cgl T}) \mid \cg U_{\cgl S}. \label{five}
\end{equation}
Similarly,  from  the hypothesis $\cg U_{\cgl S} \ind \cg U_{\cgl T}$ and (\ref{three}$ii$), we get
\begin{equation}
 \cg U_{\cgl S} \ind (\cg R_{\cgl T}, \cg U_{\cgl T}). \label{six}
\end{equation}
The contraction property is further used to write the conditional independences \eqref{five} and \eqref{six} in an equivalent condition
$
 (\cg R_{\cgl S}, \: \cg U_{\cgl S}) \: \ind \: (\cg R_{\cgl T}, \: \cg U_{\cgl T}),
$
which implies $\cg R_{\cgl S} \ind \cg R_{\cgl T}$. \begin{flushright}
$\Box$
\end{flushright}

\subsection{A marginal parameterization}\label{Sec.parameterization}

The mixture (\ref{mixt}) and Theorem \ref{teo1} characterize the probability function of the observable variables in terms of the uncertainty distributions $g_i(r_i)$, $i=1,2,...,v,$ and the marginal distributions of $p(\b r| \b u^*)$. However, we need an explicit parameterization   to tackle identifiability issues and  parameter non-redundancy, to model covariate effects and to compute Maximum Likelihood (ML) estimates. The following theorem introduces a marginal parameterization \citep{bergsma2002, bartolucci2007, colombi2014hmmm} which is extremely convenient to deal with these problems.
\begin{Theorem}\label{teo4} Under A1, A2 and A3, the probability function $p(\b r)$ can be parameterized by the following marginal interactions
	\begin{itemize}
		\item[i)] the Glonek-McCullagh interactions defined on the marginal distributions of $\pi(\b u),$
		\item[ii)] the vectors of logits $\b \eta^{\{R_i\}}$ of the  probability functions $g_i(r_i)$, $i \in \cgl V$ of the uncertain responses,
		\item[iii)]the vectors of log odds ratios $\b \eta^{\{R_i, U_i\}}$, given by the difference between the vector of logits $\b \eta^{\{R_i\mid \b u^*\}}$ of  $p(\b r_{\{i\}}|\b u^*)$ and
		$\b \eta^{\{R_i\}}$, $i \in \cgl V$,
		\item[iv)] the vectors of log odds ratios $\b \eta^{\{R_i, R_j\mid \b u^*\}}$ computed on the bivariate distributions $p(\b r_{\{i,j\}}|\b u^*)$, $i\in \cgl V, j \in \cgl V, i\neq j$.
	\end{itemize}
\end{Theorem}

\emph{Proof.} The interactions $i)$ and the logits $ii)$ parameterize  the probabilities $ \pi(\b u)$ and  $g_i(r_i)$, $i \in \cgl V$, respectively. The vector of logits
$\b \eta^{\{R_i\}}+\b \eta^{\{R_i, U_i\}}$ parametrizes the univariate marginal probability functions of $p(\b r |\b u^*)$ and, together with the log odds ratios $iv)$, parameterize the bivariate marginal probability functions.
Now, as a consequence of Assumption A3, all the Glonek-McCullagh interactions among more than two variables are set to zero. Therefore, the parameters $\b \eta^{\{R_i\}}+\b \eta^{\{R_i, U_i\}}$ and $\b \eta^{\{R_i, R_j\mid \b u^*\}}$ are sufficient to parameterize $p(\b r |\b u^*)$. Then, the proof follows
from \eqref{mixt} and Theorem \ref{teo1}. \hfill $\Box$\\
 
The uncertainty distributions $g_i(r_i)$, $i \in \cgl V$, mentioned at point \emph{ii)} of Theorem \ref{teo4} and in Section \ref{Sec.bivariate} for the bivariate case, can be chosen among distributions not depending on any unknown parameter. 
As an alternative, when possible, one can choose more flexible uncertainty distributions depending on few unknown parameters. 
As a possible choice, we propose the Local (Global) Reshaped Parabolic distribution. This is a function of the local  (global) odds of a Parabolic distribution to the power of a parameter $\phi$, which acts as a shape parameter. 
High values of the shape parameter correspond to the case where the uncertain response is focused on middle categories, while low values coincide with uncertainty focused on extreme categories.
Consequently,  the  Reshaped Parabolic distribution can model different response styles as resoluteness in the extremes or middle responses \citep[see][among others]{baumgartner2001} in the process of answering.
Details are given in the Appendix B. 

Notice that when the uncertainty distribution is a Local (Global) Reshaped Parabolic probability function and $\b \eta^{\{R_i\}}$ is a vector of local  (global) logits, then $\b \eta^{\{R_i\}}=\phi_i\b l_i$, where $\b l_i$ are vectors of known constants, $i=1,2,...,v$. This is a very useful property of this distribution, having  the vector of logits linearly depending  on a single parameter, for each variable.

Under multinomial sampling, the ML estimates  of the  parameters can be computed
by maximizing the log-likelihood function via the Fisher scoring or BFGS algorithm and it is not necessary to resort to the slower EM algorithm, commonly used with mixture models. Details on ML estimates are reported in Appendix A. An R-function that maximizes the  log-likelihood function  and computes the ML estimates with their estimated standard errors is available from the authors. The function relies on the package \texttt{hmmm} \citep{colombi2014hmmm}.

\subsection{Identifiability conditions }\label{Sec.ident}
In this section, we discuss some necessary conditions for the identifiability of the HMMLU, besides the basic requirement $p<(m-1)$ on the number $p$ of parameters, which is usually satisfied under Assumption A3.

Mixtures like \eqref{mixt} are unidentified  when some parameter values make indistinguishable two components of the mixture \citep[Section 1.3]{Silvia2006}. The next theorem shows that requiring $\b \eta^{\{R_i, U_i\}}\neq 0, \: i=1,2,...,v,$ is  necessary to avoid this problem of non-identifiability.
\begin{Theorem} \label{teo4bis}
If there exists an $i \in v$ such that  $\b \eta^{\{R_i,U_i\}}=0, \: i=1,2, \dots, v,$ then the HMMLU is not identifiable.
\end{Theorem}
\emph{Proof.} If $\b \eta^{\{R_i,U_i\}}=0$, the vector of logits,  defined on the marginal distributions  of  $R_i$, given $U_i=1$, is equal to  $\b \eta^{\{R_i\}}$ which is the vector of logits of the distributions of $R_i$, given $U_i=0.$
In this case, the component $p(\b r \: | \: \b u)$, where $\cgl V(\b u)=\cgl V,$ is indistinguishable from the component related to the uncertain configuration where only $u_i$ is equal to one. Thus, there are infinite $ \pi(\b u)$ corresponding to the same marginal probability function $\b p(\b r)$. \hfill $\Box$

\vspace{0.3cm}


Another identifiability issue is due to the case of
a   null $ \pi(\b u)$ 
that makes $ p(\b r)$ not depending on the parameters of the component with null weight.
This problem  is usually avoided by assuming that the weights $ \pi(\b u)$ of the mixture are strictly positive. The next theorem shows that  in our case a less stringent condition is sufficient.

\begin{Theorem}\label{teo5}
Suppose A1, A2 and A3 hold. If for every couple of observable variables $R_i,R_j$, there exists an uncertain configuration $\b u$ such that $\pi(\b u)>0$ and $\{i,j\} \subset \cgl V \setminus \cgl V(\b u)$ and if for every $R_i$ there exists
a $\b u$ such that $\pi(\b u)>0$ and $i \in \cgl V(\b u)$,
then $ p(\b r)$ is a function of the parameters listed in $ii)$, $iii)$ and $iv)$ of Theorem \ref{teo4}.
\end{Theorem}
\emph{Proof.} This follows because  the parameters, listed in $iii)$ and $iv)$ of Theorem \ref{teo4}, are needed to parameterize the  distributions of the responses in the configurations with $\pi(\b u)>0$. The second condition of the theorem assures the dependence of $p (\b r)$ on the  parameters $\b \eta^{\{R_i\}}$, $i=1,2,...,v$. \hfill $\Box$

\vspace{0.3cm}
Notice that the condition $\pi(\b u)>0$ when $\b u = \b 1_v$ and $\b 0_v$ is sufficient for the conclusion of Theorem \ref{teo5}. 
Moreover, remind  that the condition that $ p(\b r)$ is a function of all  the  parameters listed in $ii)$, $iii)$ and $iv)$ of Theorem \ref{teo4} is only necessary for identifiability.
In  Appendix A, a local identifiability condition, based on the rank of the Fisher matrix, is discussed.

A further necessary condition for identifiability concerns the case of respondents  grouped into $H$ strata, corresponding to distinct configurations of some discrete observable covariates.
Notice that a suffix $h$, $h=1,2,...,H$, is added to the vectors of interactions listed in Theorem \ref{teo4}, $\b \eta_h^{\{R_i\}}$, $\b \eta_h^{\{R_i, U_i\}}$, $\b \eta^{\{R_i, R_j\mid \b u^*\}}$, and to the shape parameters, $\phi_{ih}$, of the Reshaped Parabolic distributions, when these distributions are assumed for the uncertainty component.
%
%
%
If Reshaped Parabolic distributions, or whatsoever distribution depending on a single parameter, model uncertain responses, the mixture components in \eqref{mixt} are parameterized by $Hv$ shape parameters $\phi_{ih}$,  $H\sum_{i=1}^v(m_i-1)$ elements of the vectors $\b \eta_h^{\{R_i,U_i\}}$ and $H\sum_{i=1}^{v}\sum_{j>i}^{v}(m_i-1)(m_j-1)$ log odds ratios, entries of the vectors $\b \eta^{\{R_i, R_j\mid \b u^*\}}$. Consequently,  a necessary condition of identifiability is that the number of parameters $p$ is smaller than the number of free frequencies, that is it must be
$$(2^v-1)+v+\sum_{i=1}^v(m_i-1)+\sum_{i=1}^{v}\sum_{j>i}^{v}(m_i-1)(m_j-1)\leq (m-1).$$
Section \ref{Sec.bivariate} illustrates that in the bivariate case this condition is violated also when the shape parameters are null (Uniform distribution) or the uncertain distributions do not depend on unknown parameters. Therefore, restrictions on the dependence of
$\b \eta_h^{\{R_i,U_i\}}$ and $\b \eta^{\{R_i, R_j\mid \b u^*\}}$ on covariates, defining  \emph{H} strata, are needed.

When $v\geq 3$, the above necessary condition of identifiability is usually satisfied but modelling parsimoniously the dependence of
$\b \eta_h^{\{R_i,U_i\}}$ and $\b \eta_h^{\{R_i, R_j\mid \b u^*\}}$ on the covariates remains convenient, at least for simplifying the interpretability of the model.
The shape parameters of the Rehaped Parabolic uncertain distributions may be assumed to be covariate-invariant ($\phi_{ih}=\phi_{i}, \: i=1,2,...v, \: h=1,2,...,H$) to reduce the number of parameters. In addition, linear models can be adopted for taking into account the dependence of $\b \eta_h^{\{R_i,U_i\}}$ on covariates. For example, if the strata are described by a categorical variable with $H$ categories, the  model with  parallel  effect of the covariate, on the elements of the vectors $\b \eta_h^{\{R_i,U_i\}}$
$$ \eta_h^{\{R_i,U_i\}}(i_j) = \beta_i(i_j) + \beta_{ih}, \quad h=1,2,\dots, H, \: \:   j = 1, \dots, m_i-1, \: i = 1,2,\dots, v, $$
with $\beta_{i1} = 0$, reduces the number of parameters $\eta_h^{\{R_i,U_i\}}(i_j)$ from $H\sum_{i=1}^v(m_i-1)$ to $\sum_{i=1}^v(m_i-1)+(H-1)v.$
A further simplification comes by assuming independence between the observable variables that corresponds to zero constraints on the log odds ratios  $\b \eta_h^{\{R_i, R_j\mid \b u^*\}}$.

\section{A simulation study}\label{MonteCarlo}

To illustrate the performance of the proposed model and the consequences of ignoring uncertainty in the responses, we conducted a Monte Carlo simulation study from three different scenarios.
For each scenario, we generated 100 random samples from the distribution \eqref{mistura}  proposed in Section \ref{Sec.bivariate}.
On each sample, we fitted the correct model using the parameterization presented in Theorem \ref{teo4}, for the bivariate case. Moreover, we fitted the marginal model that ignores the existence  of uncertainty in responding, wrongly assuming $\pi_{11}=1$.

In each scenario, it is $m_l=4$, $\pi_l= P(U_l=1) = 0.7$ for $l=1,2$, and no covariate is included. The uncertainty distribution is assumed Uniform.
The remaining parameter settings, specific for the three scenarios, are as follows.

\noindent \emph{Scenario A}: We set the log odds ratio for the latent variables $U_1$ and $U_2$ at 2. The marginal distribution of $R_l \mid U_l = 1$ is $(0.1, 0.2, 0.3, 0.4)$ for $l=1,2$. The association for the observable variables is modelled with all the local log odds ratios $\eta^{\{R_1, R_2\mid \b u^*\}}(i_1, i_2)= \eta^{\{R_1,R_2\mid \b u^*\}}=3$, for $i_1, i_2 = 1,2,3$.

\noindent \emph{Scenario B}: The setup is similar to Scenario A except that $U_1$ and $U_2$ are independent and the marginal distribution of $R_2 \mid U_2 = 1$ is $(0.4, 0.3, 0.2, 0.1)$.

\noindent \emph{Scenario C}: The same as in Scenario B, but the marginal distribution of $R_1 \mid U_1 = 1$ is $(0.4, 0.1, 0.1, 0.4)$ and of $R_2 \mid U_2 = 1$ is $(0.1, 0.4, 0.4, 0.1)$.

\begin{table}
	\centering
	\captionsetup{font=footnotesize,width=0.8\textwidth}
	\caption{Monte Carlo averages and standard deviations of parameter estimates  under the correct model specification and ignoring  uncertainty, with sample size $n=1\,000$} \label{SimTab1000}
	\renewcommand{\arraystretch}{1}
	\setlength{\tabcolsep}{1.4 mm}
	\resizebox*{0.85\textwidth}{!}{\begin{tabular}{lcccccccccc}
			\hline \\
			& $\eta_h^{\{R_1,U_1\}}(1)$ & $\eta_h^{\{R_1,U_1\}}(2)$ & $\eta_h^{\{R_1,U_1\}}(3)$ & $\eta_h^{\{R_2,U_2\}}(1)$ & $\eta_h^{\{R_2,U_2\}}(2)$ & $\eta_h^{\{R_2,U_2\}}(3)$ & $ \eta^{\{R_1, R_2\mid \b u^*\}}$  & $ \eta^{\{U_1\}}$ & $\eta^{\{U_2\}}$ &$\eta^{\{U_1,U_2\}}$\\
			\hline \hline
			\\
			& \multicolumn{10}{c}{\emph{Scenario A}}\\ \cline{2-11}
			True & 0.69 & 0.41 & 0.29 & 0.69 & 0.41 & 0.29 & 3.00  & 0.85 & 0.85 & 2.00 \\
			\\
			\multicolumn{2}{l}{\underline{Correct model specification}}\\
			MC Average & 0.74 & 0.40 & 0.29 & 0.72 & 0.41 & 0.29 & 3.09 & 0.86 & 0.88 & 2.19\\
			MC sd & 0.20 & 0.13 & 0.09 & 0.20 & 0.14 & 0.11 & 0.27 & 0.31 & 0.35 & 1.60\\
			\multicolumn{2}{l}{\underline{Ignoring uncertainty}}\\
			MC Average & 0.40 & 0.28 & 0.22 & 0.40 & 0.28 & 0.22 & 0.50 \\
			MC sd & 0.11 & 0.08 & 0.07 & 0.11 & 0.09 & 0.08 & 0.04 \\
			\\
			
			& \multicolumn{10}{c}{\emph{Scenario B}}\\ \cline{2-11}
			
			True  & 0.69 & 0.41 & 0.29 & -0.29 & -0.41 & -0.69 & 3.00 & 0.85 & 0.85 & 0.00\\
			\\
			\multicolumn{2}{l}{\underline{Correct model specification}}\\
			
			MC Average & 0.72 & 0.39 & 0.30 & -0.30 & -0.40 & -0.68 & 2.59 & 0.85 & 0.90 & 0.34\\
			MC sd & 0.21 & 0.13 & 0.10 & 0.11 & 0.12 & 0.20 & 0.28 & 0.32 & 0.37 & 1.55\\
			\multicolumn{2}{l}{\underline{Ignoring uncertainty}}\\
			MC Average  & 0.40 & 0.27 & 0.23 & -0.23 & -0.28 & -0.38 & 0.34\\
			MC sd  & 0.11 & 0.09 & 0.08 & 0.08 & 0.08 & 0.10 & 0.04\\
			
			\\
			& \multicolumn{10}{c}{\emph{Scenario C}}\\ \cline{2-11}
			
			True  & -1.39 & -0.00 & 1.39 & 1.39 & 0.00 & -1.39 & 3.00 & 0.85 & 0.85 & 0.00\\
			\\
			\multicolumn{2}{l}{\underline{Correct model specification}}\\
			
			MC Average & -1.37 & 0.01 & 1.37 & 1.43 & 0.01 & -1.40 & 3.19 & 0.95 & 0.85 & 0.16\\
			MC sd & 0.26 & 0.23 & 0.23 & 0.21 & 0.10 & 0.18 & 0.97 & 0.41 & 0.29 & 1.35\\
			\multicolumn{2}{l}{\underline{Ignoring uncertainty}}\\
			MC Average  & -0.89 & 0.00 & 0.89 & 0.91 & 0.00 & -0.89 & 0.38\\
			MC sd  & 0.09 & 0.12 & 0.10 & 0.11 & 0.08 & 0.10 & 0.04\\
			
			\hline
	\end{tabular}}
\end{table}

\begin{table}
	\centering
	\captionsetup{font=footnotesize,width=0.8\textwidth}
	\caption{Monte Carlo averages and standard deviations of parameter estimates under the correct model specification and ignoring  uncertainty, with sample size $n=10\,000$}\label{SimTab10000}
	\renewcommand{\arraystretch}{1.1}
	\setlength{\tabcolsep}{1.8 mm}
	\resizebox*{0.85\textwidth}{!}{\begin{tabular}{lcccccccccc}
			
			\hline \\
			& $\eta_h^{\{R_1,U_1\}}(1)$ & $\eta_h^{\{R_1,U_1\}}(2)$ & $\eta_h^{\{R_1,U_1\}}(3)$ & $\eta_h^{\{R_2,U_2\}}(1)$ & $\eta_h^{\{R_2,U_2\}}(2)$ & $\eta_h^{\{R_2,U_2\}}(3)$ & $ \eta^{\{R_1, R_2\mid \b u^*\}}$  & $ \eta^{\{U_1\}}$ & $\eta^{\{U_2\}}$ &$\eta^{\{U_1,U_2\}}$\\
			\hline \hline
			\\
			& \multicolumn{10}{c}{\emph{Scenario A}}\\ \cline{2-11}
			True & 0.69 & 0.41 & 0.29 & 0.69 & 0.41 & 0.29 & 3.00  & 0.85 & 0.85 & 2.00 \\
			\\
			\multicolumn{2}{l}{\underline{Correct model specification}}\\
			MC Average & 0.69 & 0.40 & 0.29 & 0.69 & 0.40 & 0.29 & 3.00  & 0.88 & 0.88 & 1.83\\
			MC sd & 0.06 & 0.04 & 0.03 & 0.05 & 0.04 & 0.03 & 0.12 & 0.10 & 0.09 & 0.52\\
			\multicolumn{2}{l}{\underline{Ignoring uncertainty}}\\
			MC Average & 0.39 & 0.28 & 0.22 & 0.40 & 0.28 & 0.22 & 0.49 \\
			MC sd & 0.03 & 0.03 & 0.03 & 0.03 & 0.03 & 0.03 & 0.01 \\
			\\

			& \multicolumn{10}{c}{\emph{Scenario B}}\\ \cline{2-11}
			True & 0.69 & 0.41 & 0.29 & -0.29 & -0.41 & -0.69 & 3.00 & 0.85 & 0.85 & 0.00 \\
			\\
			\multicolumn{2}{l}{\underline{Correct model specification}}\\
			MC Average & 0.69 & 0.41 & 0.29 & -0.28 & -0.40 & -0.70 & 2.94 & 0.84 & 0.85 & 0.08\\
			MC sd & 0.06 & 0.04 & 0.04 & 0.03 & 0.04 & 0.06 & 0.17 & 0.11 & 0.10 & 0.43\\
			\multicolumn{2}{l}{\underline{Ignoring uncertainty}}\\
			MC Average & 0.39 & 0.28 & 0.22 & -0.22 & -0.28 & -0.40 & 0.34\\
			MC sd & 0.04 & 0.03 & 0.03 & 0.02 & 0.03 & 0.03 & 0.01\\
			\\
			& \multicolumn{10}{c}{\emph{Scenario C}}\\ \cline{2-11}
			True &  -1.39 & -0.00 & 1.39 & 1.39 & 0.00 & -1.39 & 3.00 & 0.85 & 0.85 & 0.00 \\
			\\
			\multicolumn{2}{l}{\underline{Correct model specification}}\\
			MC Average & -1.39 & -0.00 & 1.39 & 1.39 & -0.00 & -1.40 & 3.12 & 0.88 & 0.84 & -0.00\\
			MC sd & 0.11 & 0.08 & 0.10 & 0.07 & 0.03 & 0.06 & 0.63 & 0.17 & 0.09 & 0.32\\
			\multicolumn{2}{l}{\underline{Ignoring uncertainty}}\\
			MC Average & -0.90 & -0.00 & 0.90 & 0.90 & -0.00 & -0.90 & 0.37\\
			MC sd & 0.03 & 0.04 & 0.03 & 0.03 & 0.02 & 0.03 & 0.01\\
			\hline
	\end{tabular}}
\end{table}

\begin{figure}
	\centering
	\begin{tabular}{p{5cm} p{5cm} p{5cm}}
		\multicolumn{3}{c}{Scenario A}\\
		\includegraphics[clip,trim=2cm 1cm 1cm 1cm,width=5cm]{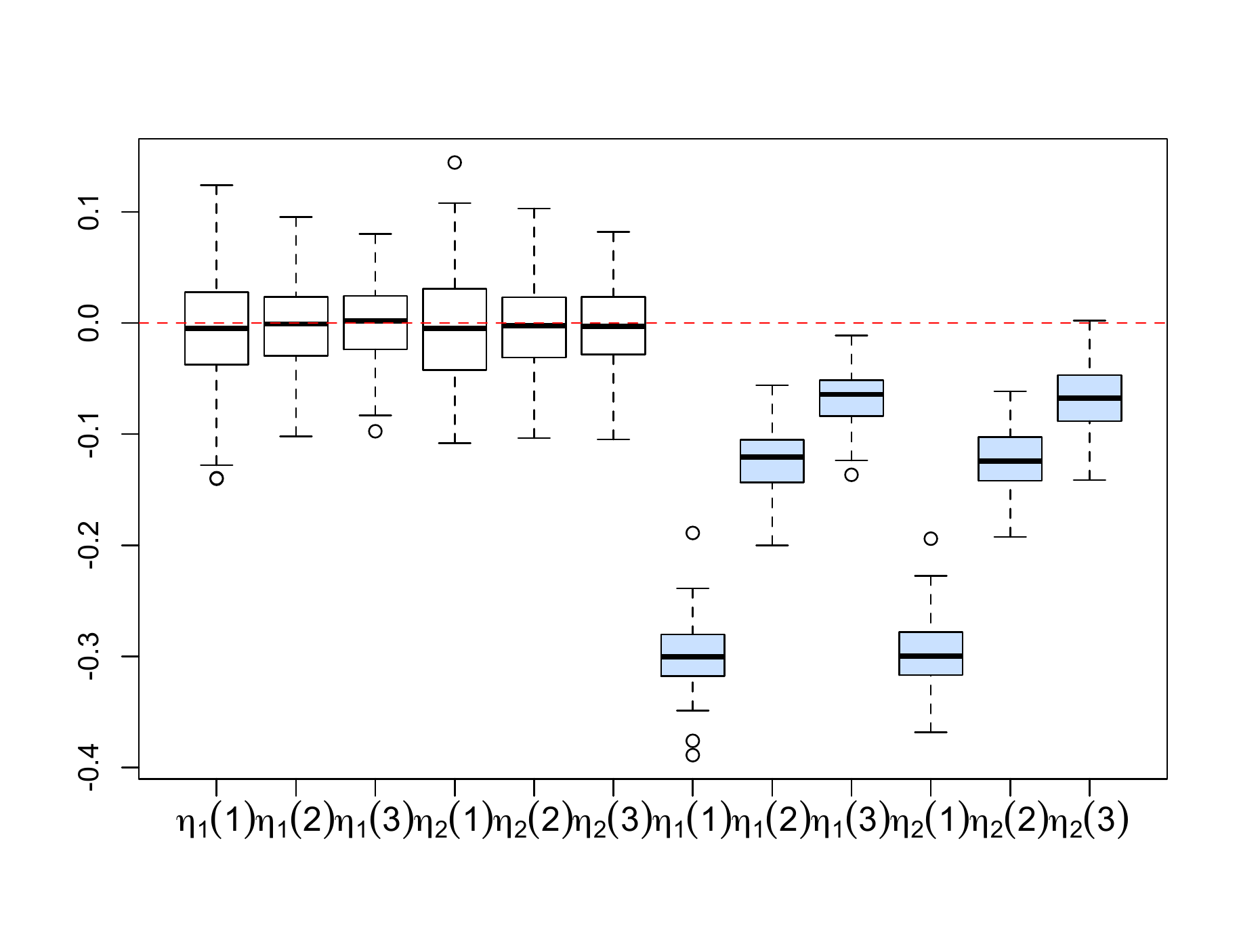} &
		\includegraphics[clip,trim=2cm 1cm 1cm 1cm,width=5cm]{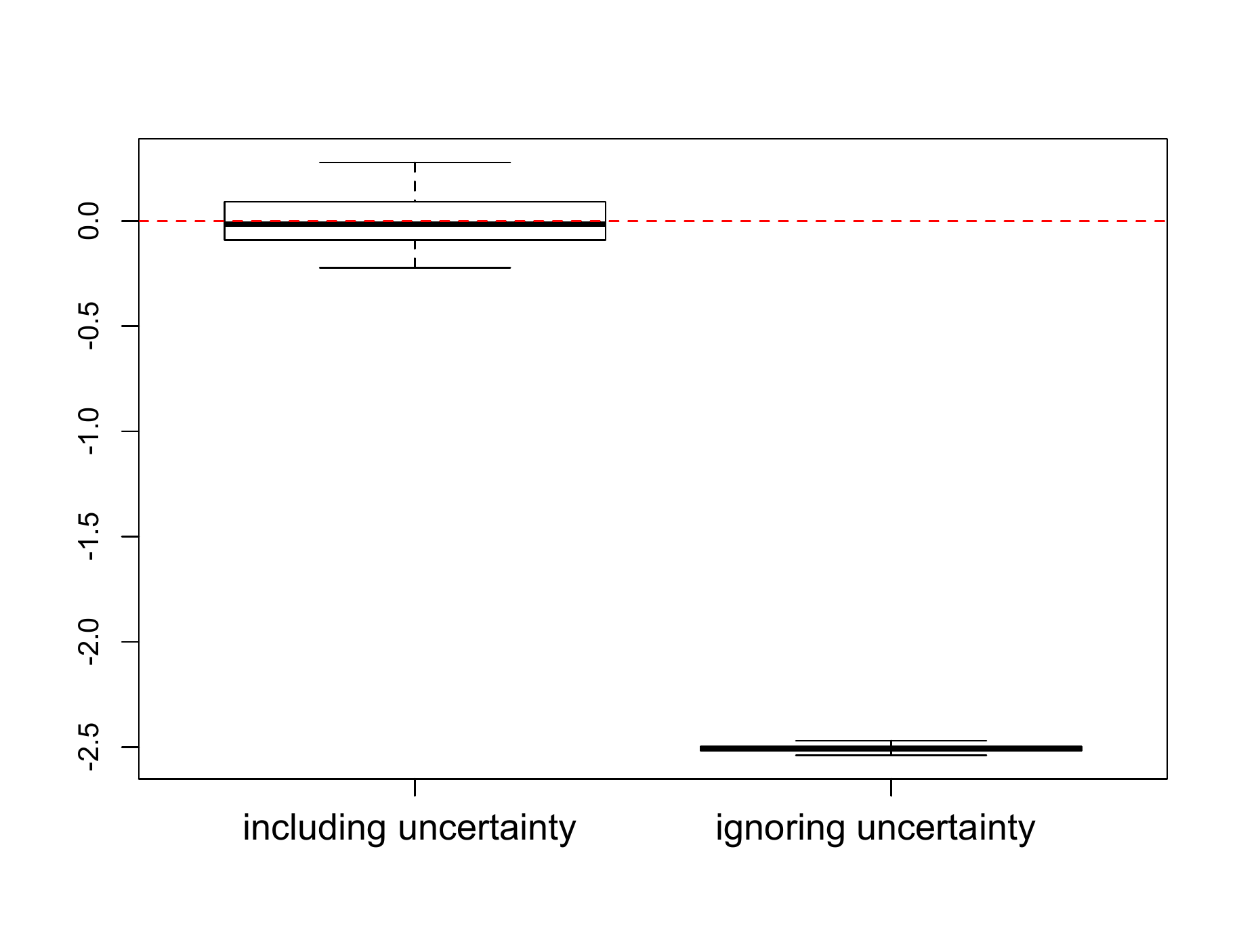}&
		\includegraphics[clip,trim=2cm 1cm 1cm 1cm,width=5cm]{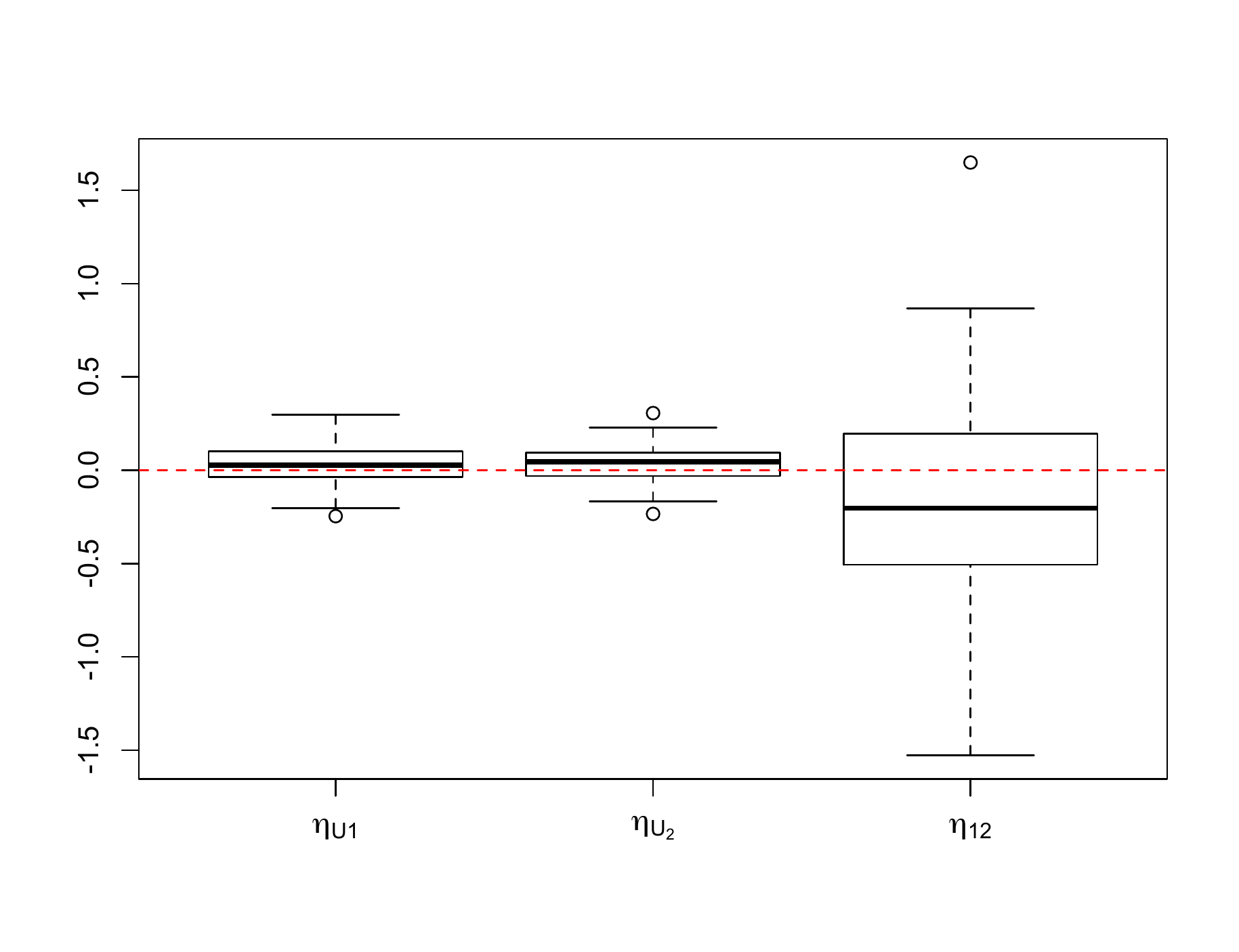}
		\\
		\multicolumn{3}{c}{Scenario B}\\
		\includegraphics[clip,trim=2cm 1cm 1cm 1cm,width=5cm]{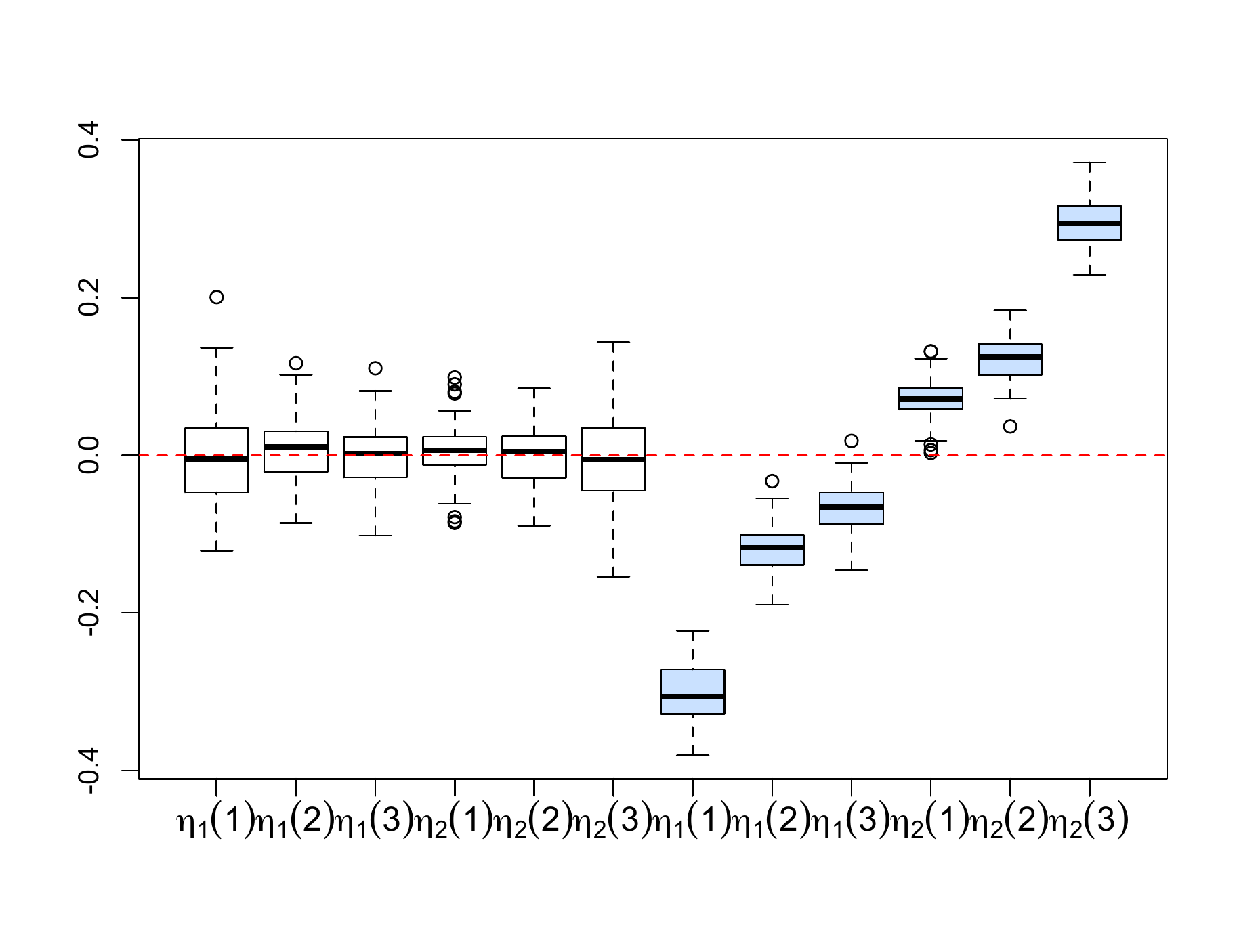} &
		\includegraphics[clip,trim=2cm 1cm 1cm 1cm,width=5cm]{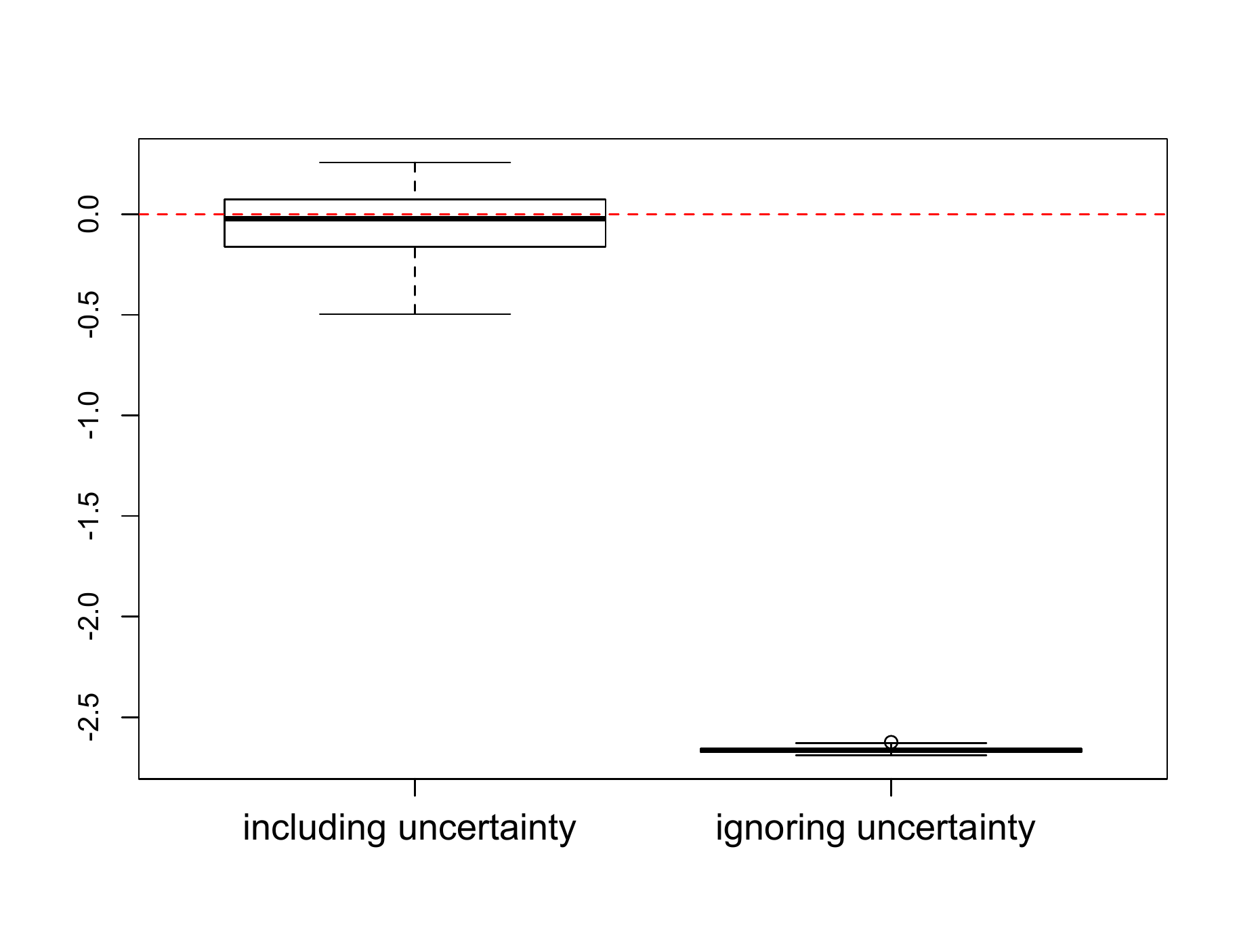}
		&\includegraphics[clip,trim=2cm 1cm 1cm 1cm,width=5cm]{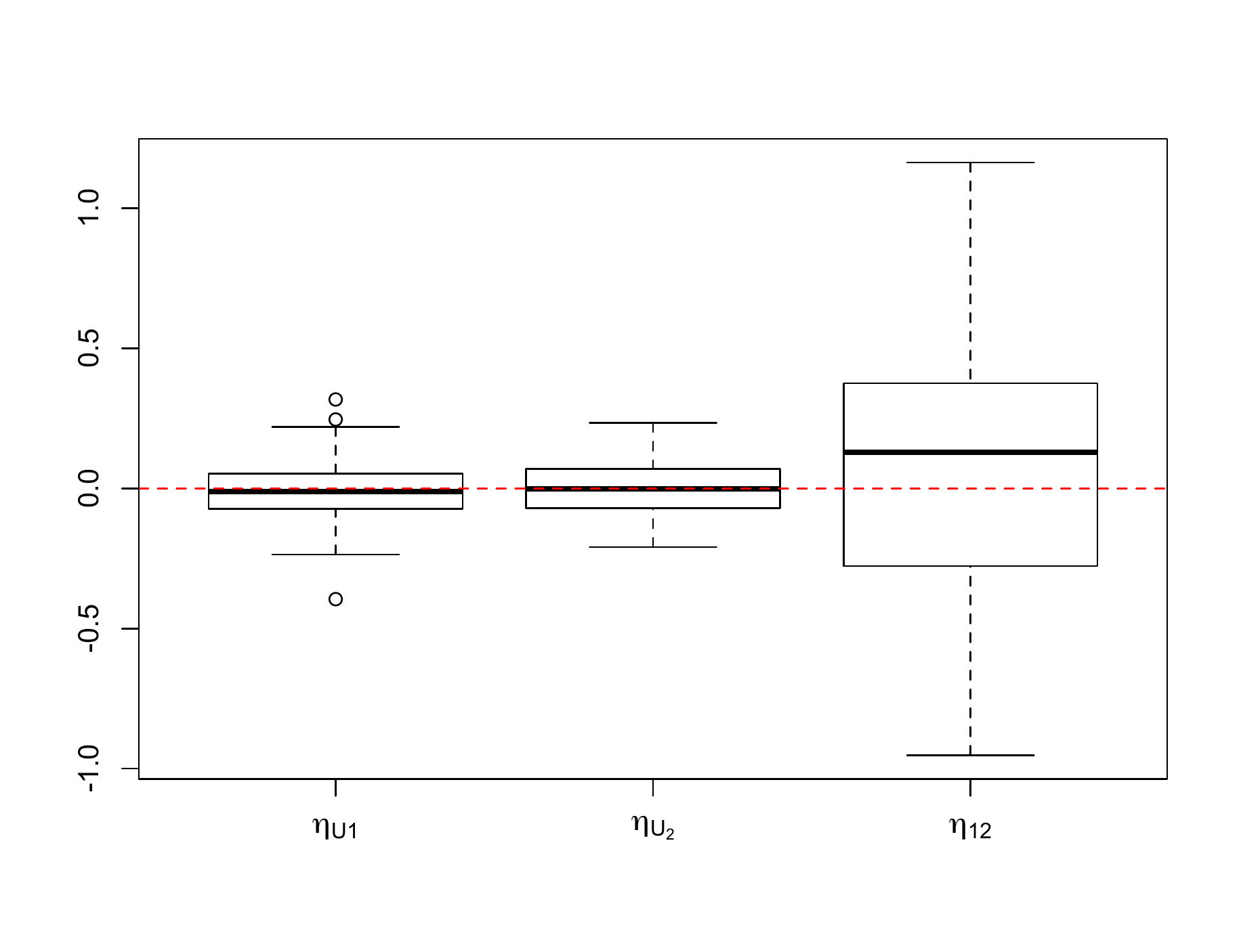}\\
		\multicolumn{3}{c}{Scenario C}\\
		\includegraphics[clip,trim=2cm 1cm 1cm 1cm,width=5cm]{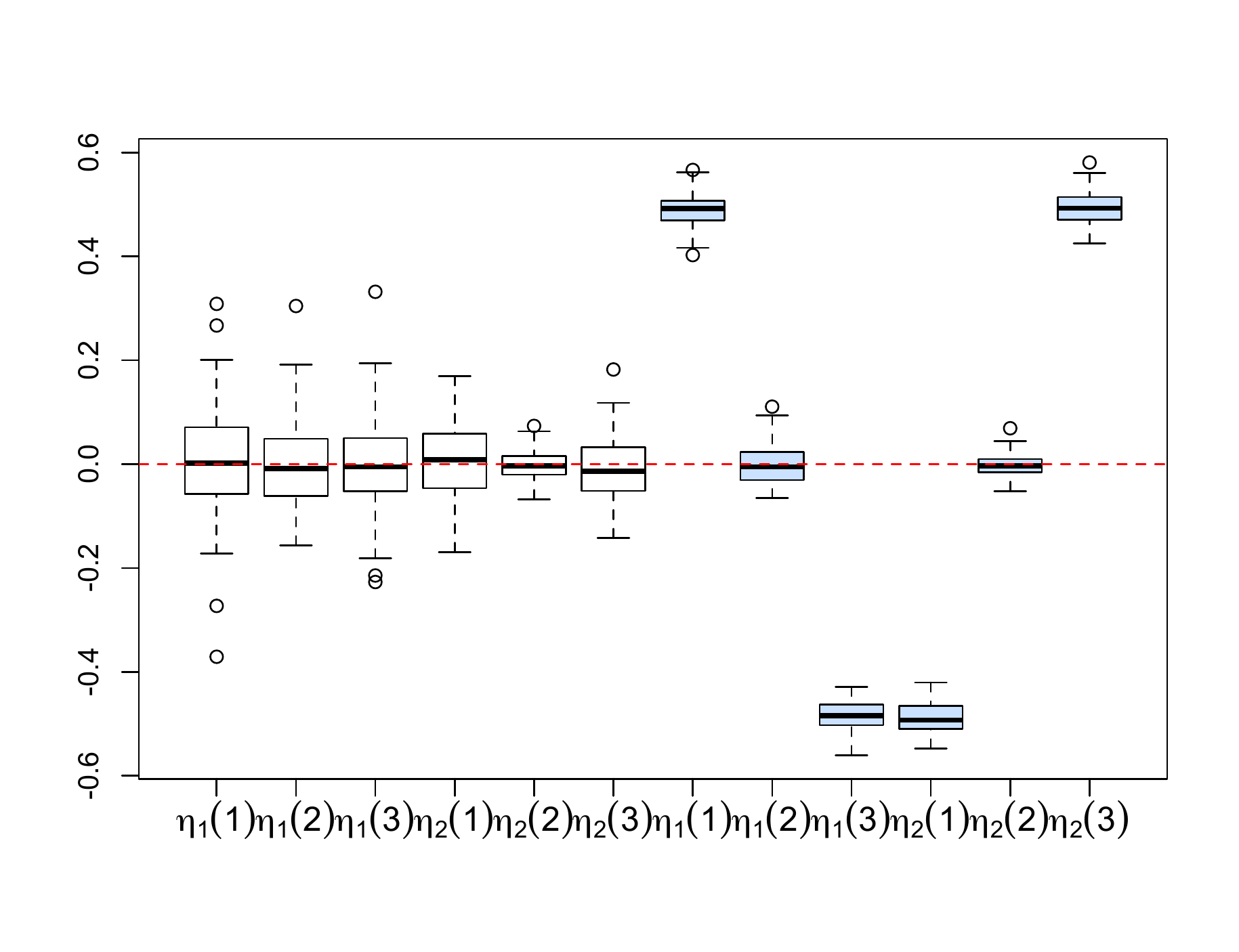} &
		\includegraphics[clip,trim=2cm 1cm 1cm 1cm,width=5cm]{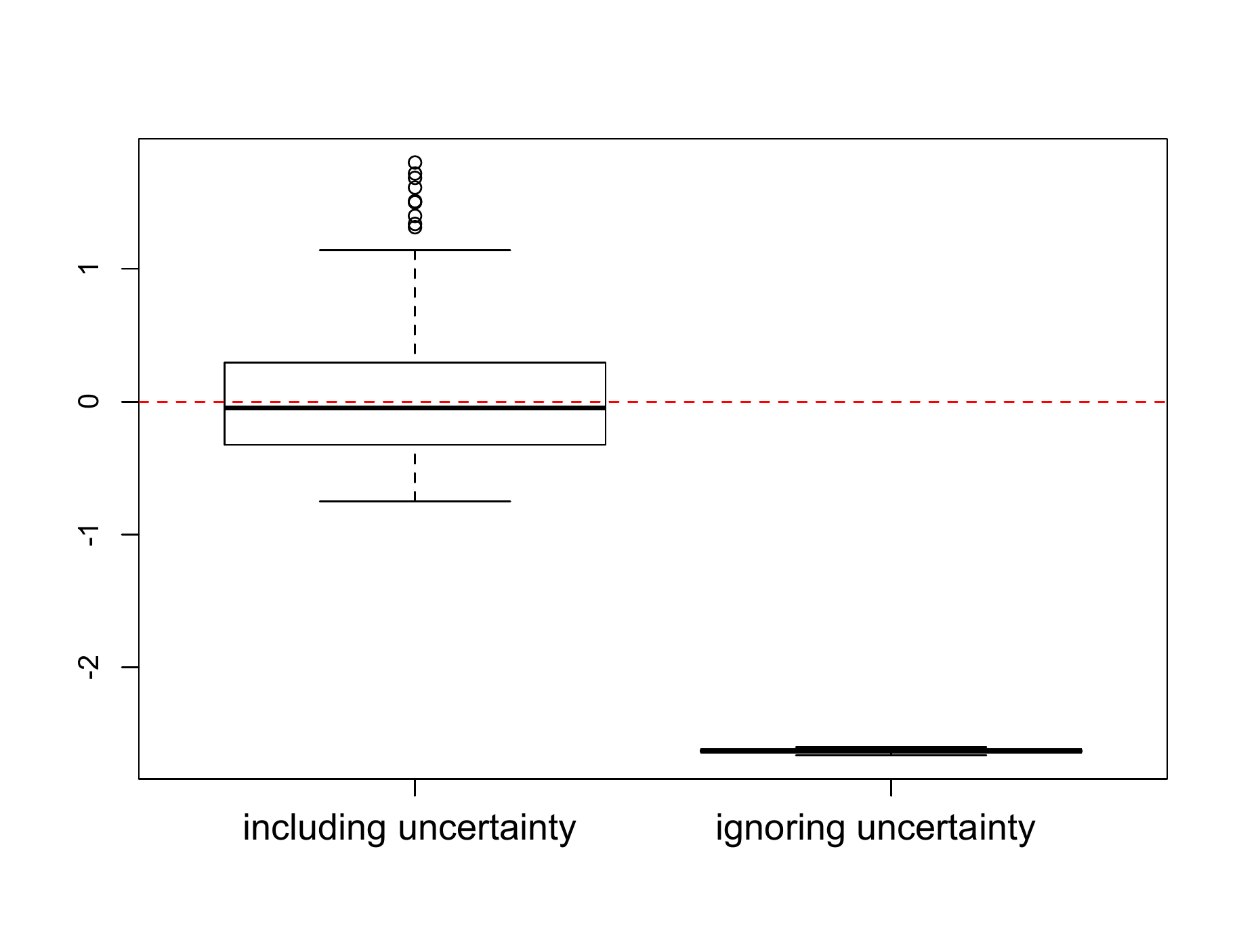}
		&\includegraphics[clip,trim=2cm 1cm 1cm 1cm,width=5cm]{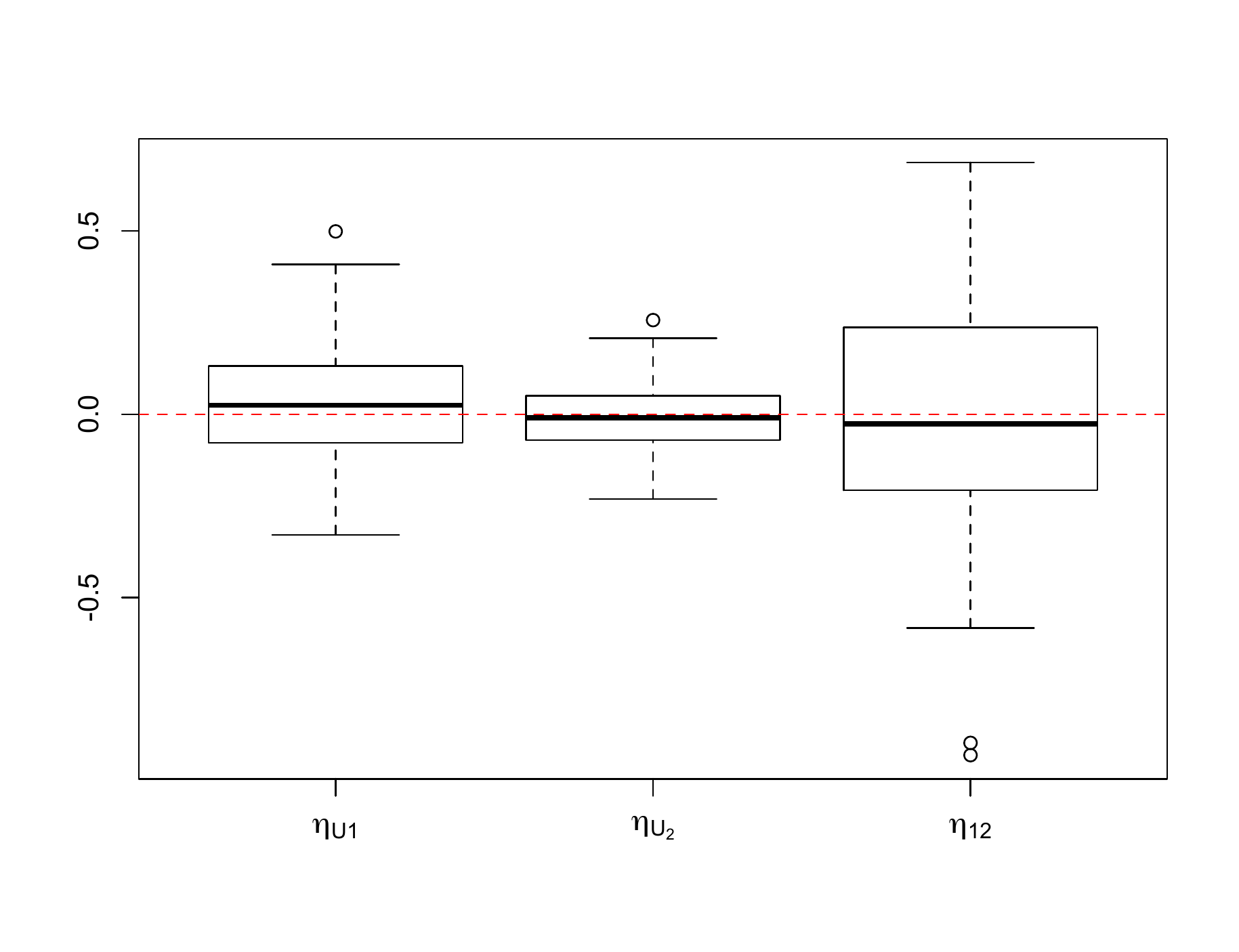}\\
	\end{tabular}
	\captionsetup{font=footnotesize,width=0.95\textwidth}\caption{Box plot of Monte Carlo errors for marginal logit (first column) and log odds ratio (second column), including (white) and ignoring (coloured) uncertainty, and latent variable parameters (third column) with $n = 10\,000$}\label{Boxplot}
\end{figure}

The Monte Carlo experiment was repeated for sample size  $n=1\,000$ and $n=10\,000$ to evaluate the asymptotic behaviour of the estimates. Summaries of the simulation results are reported in Tables \ref{SimTab1000} and \ref{SimTab10000}. 
In these tables, the  local logits and log odds ratios parameters corresponding to the three scenarios are reported in the lines labelled as \emph{True}. Moreover, the last three columns concern the Gloneck-McCullagh interactions defined at point $i)$ of Theorem \ref{teo4}.
As can be seen along the tables, the proposed estimation procedure is able to capture quite well the model parameters concerning the rating of the two items in the aware component ($U_1=U_2=1$). On the other hand, the estimates for the model parameters obtained ignoring uncertainty, well illustrate the consequences of model misspecification. These consequences are better detectable in Figure \ref{Boxplot} that presents the box plots for the Monte Carlo errors under the proposal models (white) and ignoring uncertainty (coloured).

Estimates from the model ignoring uncertainty differ substantially from the true values and underestimate or overestimate the true parameters.
As a matter of fact, ignoring uncertainty corresponds to estimating logits and log odds ratios of the mixture of four components \eqref{mistura}, when actually we are interested in the parameters of the fourth component of this mixture.
In particular, in Scenario A   the local logits, all positive, are underestimated. On the contrary, in Scenario B the negative local logits of $R_2 \mid U_2 = 1$ are overestimated.
A similar pattern can be detected in Scenario C for the positive and negative logits of the probability functions
$(0.4, 0.1, 0.1, 0.4)$ and  $(0.1, 0.4, 0.4, 0.1)$.  This is explained by the fact  that  in the marginal distribution of the observable variables, the logits shrink in absolute values because of the Uniform component. Analogously, in all the considered scenarios, the positive uniform association in the fourth component of the mixture \eqref{mistura} is underestimated if uncertainty is not taken into account.

\section{Illustrative example}\label{example}

This example concerns the perception of the quality of working life, using data from the 5$^{th}$ European Working Conditions Survey (EWCS). The survey has been carried out by  the European Foundation for the Improvement of Living and Working Conditions (Eurofound).
We focus on respondents' agreement on three statements: $R_1=$  \emph{Losejob} (I might lose my job in the next 6 months),	$R_2 =$ \emph{Wellpaid} (I am well paid for the work I do) and
$R_3 = $ \emph{Career} (My job offers good prospects for career advancement).
The responses are recoded on a 3-point scale: \emph{disagree}, \emph{neither agree nor disagree}, \emph{agree}. In addition, we consider two explanatory variables, \emph{Gender} (0 = \emph{Male}, 1 = \emph{Female}) and \emph{Country} (0 = \emph{Northern} and 1 = \emph{Southern} EU regions   according to the geographic scheme in use by the United Nations).
Table \ref{EWCS} reports the data of a sample of $3\,500$ workers derived from those available on
the Eurofound website \texttt{https://www.eurofound.europa.eu}.

\begin{table}[h!]
	\centering
	\captionsetup{font=footnotesize,width=0.8\textwidth}
	\caption{Observed joint distribution of EWCS data for \emph{Losejob}, \emph{Wellpaid} and \emph{Career}, \emph{Gender} and \emph{Country} }\label{EWCS}
	\renewcommand{\arraystretch}{0.8}
	\setlength{\tabcolsep}{2 mm}
	\resizebox*{.9\textwidth}{!}{\begin{tabular}{r||cccccccccccc}
			\emph{Gender}         & &       &                         &          & Male        &      & &   &Female             &\\
			\hline
			& &       &          \emph{Career}  &disagree  & n. agree    & agree&&disagree  & n. agree    & agree\\
			& &       &            &  &  n. disagree   & &&  &  n. disagree   & \\
			\emph{Country}& \emph{Losejob}& &\emph{Wellpaid} &&&&&&&\\
			\hline
			& &       & &&&& &&&\\
			Northern &disagree&       &disagree         &      136&  41&  26&& 179&  62&  34\\
			regions& &       &n. agree n. disagree         &      121&  94&  84&& 129&  76&  62\\
			of EU & &       &agree         &      116&  87& 227&& 89&  57& 173\\
			& &       & &&&& &&&\\
			&n. agree n. disagree &       &disagree         &       45&  10&   7&& 30&  14&   6\\
			& &       &n. agree n. disagree         &       21&  40&  30&& 32&  20&  15\\
			& &       &agree         &       13&  19&  25&&  5&   8&  25\\
			& &       & &&&& &&&\\
			&agree&       &disagree         &       76&   7&  11&&  60&   9&  13\\
			& &       &n. agree n. disagree          &       36&  21&   4&&  13&   9&   6\\
			& &       &agree         &       18&   9&  12&&  10&  10&  18\\
			& &       & &&&& &&&\\
			Southern&disagree&      &disagree         &       30&   7&  17&&  39&  11&  19\\
			regions & &      &n. agree n. disagree          &       33&  31&  26&&  24&  20&  22\\
			of EU   & &      &agree         &       49&  64& 137&&  41&  36&  97\\
			& &      & &&&& &&&\\
			&n. agree n. disagree &       &disagree         &        9&   3&   2&& 7&   4&   3\\
			& &       &n. agree n. disagree          &       11&  10&   4&&  5&   3&   2\\
			& &       &agree         &        6&  14&  22&&  8&   8&  12\\
			& &       & &&&& &&&\\
			&agree&       &disagree         &       15&   1&   6&&  18&   2&   5\\
			& &       &n. agree n. disagree          &        8&  15&   4&&  10&  10&   5\\
			& &       &agree         &       12&  10&  21&&   9&   7&   6\\
			& &       & &&&& &&&\\
			\hline
	\end{tabular}}
\end{table}

It is reasonable to assume that not all the respondents have been able to allocate their perceptions exactly into a category when requested to evaluate personal satisfactions and worries on their work. Hence, the observed responses could have been contaminated by a certain amount of uncertain answers.
The aim of this illustrative example is to show how an HMMLU adequately takes into account for uncertainty in the responses, detects which one is perceived with more/less uncertainty, and if the proportion of uncertain answers changes with the individual characteristics.
The model can also describe the association between the aware responses and their dependence on the respondent's features, separately from the uncertain answers.

With this intent, we specify several models, with different hypotheses about the association  and/or the  dependence on covariates  \emph{Gender} and \emph{Country}.
In each model, we  adopt Local Reshaped Parabolic distributions with shape parameters independent of the explanatory variables for the uncertain responses and we use local  logits and local odds ratios to parameterize the components in \eqref{mixt}.
Table \ref{tabeuro} summarizes the fitting of some of these models.
\begin{table}[h!]
 \captionsetup{font=footnotesize,width=0.9\textwidth}
 \caption{\label{tabeuro}{Hypotheses specifying the models, log-likelihood values ($\ell$), number of parameters (\emph{n.par.}), models compared via likelihood ratio tests (\emph{LRT}) and the corresponding $p$-values
}}
 \renewcommand{\arraystretch}{0.8}
 \setlength{\tabcolsep}{2.4 mm}
 \centering  \resizebox*{0.8\textwidth}{!}{\begin{tabular}{cccccccc}
 \hline  &   &    &    &   &     &  \\
 \emph{ Model}   & \emph{Hypotheses on}  & \emph{ Hypotheses on } &   $\ell$   & \emph{n.par.}  &  \textit{Compared} & $LRT$ $p$-value  &  \\
 & \emph{obs. responses} & \emph{latent var.} &    & & \textit{models}  \\ \hline
 &   &    &    &   &     &  \\
 \multirow{3}{*}{$\mathcal{M}_{0}$} & Unrestricted ass. & Unrestricted ass. & \multirow{3}{*}{-9849.839}  & \multirow{3}{*}{103} &     &  \\
 &  with covariates  &   with covariates  &    &   &     &   &  \\
  &  unrestricted eff.  &   unrestricted eff.  &    &   &     &   &  \\
 &   &    &    &   &     &  \\
  \multirow{3}{*}{$\mathcal{M}_{1}$} & Homogeneous ass. & Unrestricted ass. & \multirow{3}{*}{-9866.294}  & \multirow{3}{*}{67} & \multirow{3}{*}{$\mathcal{M}_1$ vs $\mathcal{M}_0$} & \multirow{3}{*}{0.6164}\\
 &  with covariates  &   with covariates  &    &   &     &   &  \\
  &  unrestricted eff.  &   unrestricted eff.  &    &   &     &   &  \\
  &   &    &    &   &     &  \\
 \multirow{3}{*}{$\mathcal{M}_{2}$} & Uniform ass. & Unrestricted ass. & \multirow{3}{*}{-9906.695}  & \multirow{3}{*}{67} & \multirow{3}{*}{$\mathcal{M}_2$ vs $\mathcal{M}_0$} & \multirow{3}{*}{0.0000}\\
 &  with covariates  &   with covariates  &    &   &     &   &  \\
   &  unrestricted eff.  &   unrestricted eff.  &    &   &     &   &  \\
  &   &    &    &   &     &  \\
 \multirow{3}{*}{$\mathcal{M}_{3}$} & Homogeneous ass. & Unrestricted ass. & \multirow{3}{*}{-9883.178}  & \multirow{3}{*}{55} & \multirow{3}{*}{$\mathcal{M}_3$ vs $\mathcal{M}_0$} & \multirow{3}{*}{0.0384}\\
&  with covariates  &   with covariates  &    &   &     &   &  \\
&  additive-parallel eff.  &   unrestricted eff.  &    &   &     &   &  \\
&   &    &    &   &     &  \\
\multirow{3}{*}{$\mathcal{M}_{4}$} & Homogeneous ass. & Unrestricted ass. & \multirow{3}{*}{-9944.525}  & \multirow{3}{*}{49} & \multirow{3}{*}{$\mathcal{M}_4$ vs $\mathcal{M}_0$} & \multirow{3}{*}{0.0000}\\
&   no covariates  &   with covariates  &    &   &     &   &  \\
&  -  &   unrestricted eff.  &    &   &     &   &  \\
  &   &    &    &   &     &  \\
\multirow{3}{*}{$\mathcal{M}_{5}$} & Homogeneous ass. & Independence & \multirow{3}{*}{-9884.186}  & \multirow{3}{*}{36} & \multirow{2}{*}{$\mathcal{M}_5$ vs $\mathcal{M}_0$} & \multirow{2}{*}{0.4198}\\
&  with covariates  &   with covariates  &    &   &     &   &  \\
&  additive-parallel eff.  &   additive-parallel eff.  &    &   & $\mathcal{M}_5$ vs $\mathcal{M}_1$    &  $0.2538$ &  \\
&   &    &    &   &     &  \\
\multirow{3}{*}{$\mathcal{M}_{6}$} & Homogeneous ass. & Independence & \multirow{3}{*}{-9900.109}  & \multirow{3}{*}{30} & \multirow{3}{*}{$\mathcal{M}_6$ vs $\mathcal{M}_5$} & \multirow{3}{*}{0.0000}\\
&  with covariates  &   no covariates  &    &   &     &   &  \\
&  additive-parallel eff.  &   -  &    &   &     &   &  \\
\hline
 \end{tabular}}
\end{table}
Among the analysed models,  $\mathcal{M}_{5}$ shows the best fit on the base of the likelihood ratio test (LRT). According to this model  the latent variables are independent  and the association among the aware responses is homogeneous  \citep[see][Section 6.7.2]{kateri2014book}.
The effect of \emph{Gender} and \emph{Country} is modelled in $\mathcal{M}_{5}$ by the linear  models  with parallel and additive effect of the covariates on the parameters $\eta_{hk}^{\{R_i,U_i\}}(i_j)$ and the logits for the latent variables $\eta_{hk}^{\{U_i\}}$
\begin{align}
\label{obslogit}
 \eta_{hk}^{\{R_i,U_i\}}(i_j)  = \beta_i(i_j) + \beta^G_{ih} + \beta^C_{ik} , \qquad &\text{ with }\beta^G_{i0} = \beta^C_{i0} = 0,\\
 \label{latlogit}
\eta_{hk}^{\{U_i\}} =  \tilde{\beta}_i + \tilde{\beta}^G_{ih} + \tilde{\beta}^C_{ik},
\qquad &\text{ with } \tilde{\beta}^G_{i0} = \tilde{\beta}^C_{i0} = 0,
\end{align}
where $h= 0, 1, \: \: k = 0, 1, \: \:   i_j = 1, 2, \: \: i = 1,2,3, \: \: j = 1,2,3$.

\begin{figure}[h!]
	\centering
	\includegraphics[width=8cm]{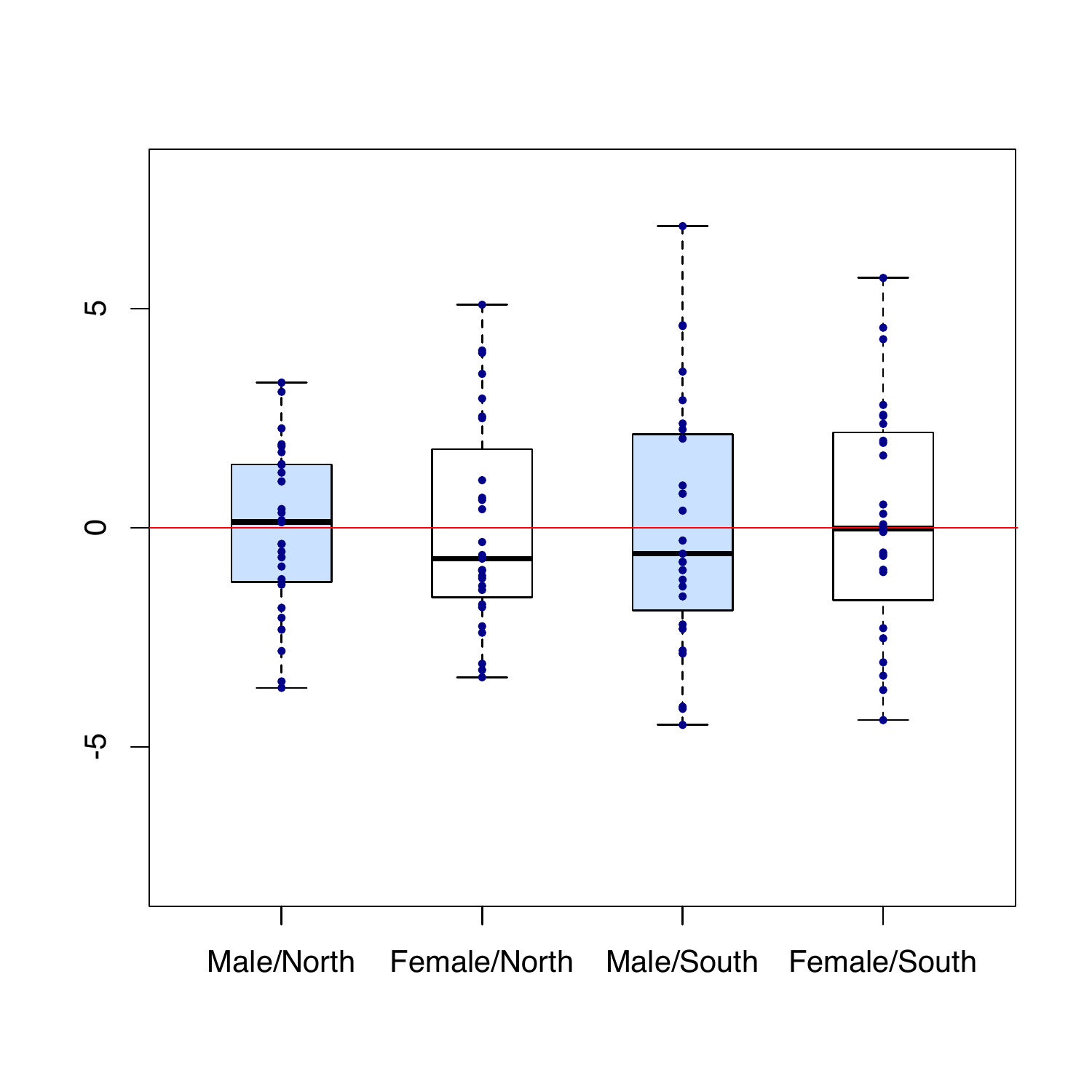}
	\vskip-.5cm\captionsetup{font=footnotesize,width=0.8\textwidth}\caption{Plot of the standardized residuals of model $\mathcal{M}_5$}\label{res}
\end{figure}

To provide an insight on the goodness-of-fit of model $\mathcal{M}_5$, in Figure \ref{res} we show  the standardized residuals, computed on joint sample frequencies and estimated probabilities, by the covariate strata.  Most of the residuals are small as over $87\%$ do not exceed the threshold 4 in absolute value showing a satisfactory fit. A more careful inspection reveals that highest residuals correspond to the stratum of Southern workers. In marginal modelling, the fitting of the univariate  marginal distributions is often the main interest and association parameters are regarded as nuisance parameters. From this point of view, the standardized marginal residuals, based on univariate sample frequencies and estimated probabilities, are  relevant. Here, the marginal residuals highlight that the marginal distributions are well fitted, except for the distribution related to \textit{Career} in Southern EU regions (see Table \ref{margres}).

\begin{table}[h!]
\captionsetup{font=footnotesize,width=0.9\textwidth}
\caption{\label{margres}{Standardized marginal residuals
}}
\renewcommand{\arraystretch}{0.9}
 \setlength{\tabcolsep}{2.4 mm}
 \centering  \resizebox*{0.6\textwidth}{!}{
\begin{tabular}{ccccc}
  \hline
 & Male North & Female North & Male South & Female South \\
  \hline
  \textit{Losejob}& & & &\\
  disagree & -0.510 & 0.331 & -0.519 & 1.533 \\
  n. agree n. disagree & 1.020 & -0.759 & 0.871 & -2.073 \\
  agree & -0.510 & 0.427 & -0.352 & 0.540 \\
  \hline
 \textit{ Wellpaid} & & & &\\
  disagree & -0.861 & 0.903 & -1.256 & 2.336 \\
  n. agree n. disagree & 2.500 & -2.545 & 1.047 & -1.678 \\
  agree & -1.639 & 1.642 & 0.209 & -0.658 \\
  \hline
\textit{ Career} & & & &\\
 disagree & 0.222 & 1.468 & -7.875 & 5.705 \\
  n. agree n. disagree & 1.065 & -1.600 & 7.082 & -7.877 \\
  agree & -1.287 & 0.133 & 0.794 & 2.173 \\
   \hline
\end{tabular}
}
\end{table}

\begin{figure} [h!]
	\centering
	\begin{tabular}{c c}
		\includegraphics[trim=0cm 0.3cm 1cm .5cm, width=6cm ]{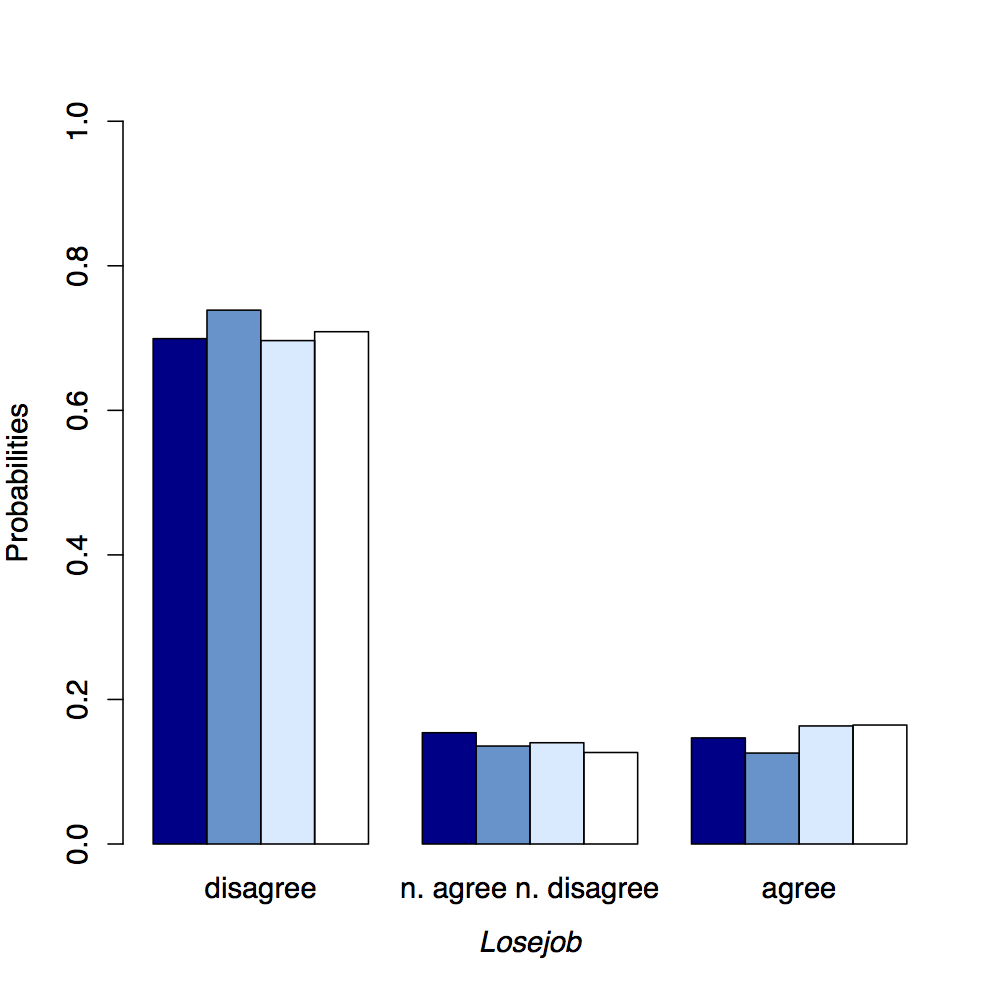} &
		\includegraphics[trim=0cm 0.3cm 1cm .5cm, width=6cm]{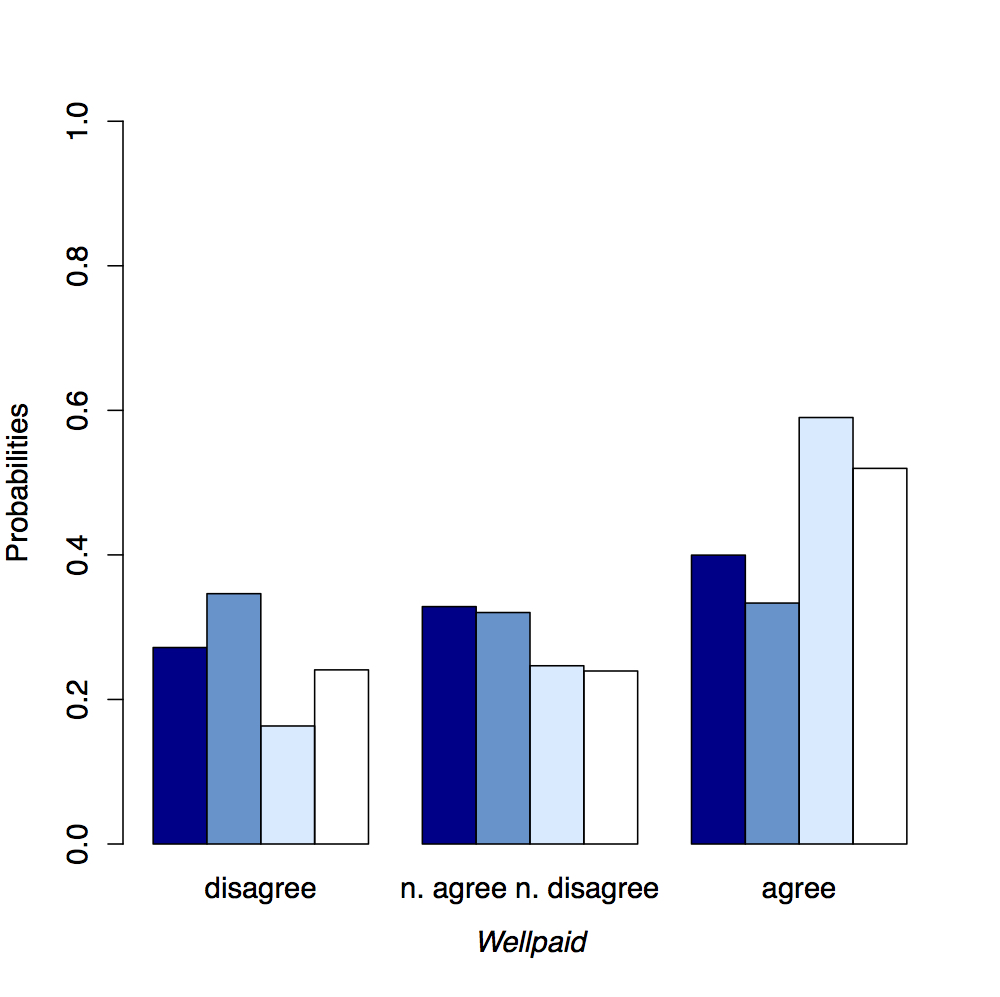} \\
		\includegraphics[trim=0cm 1cm 1cm .5cm, width=6cm]{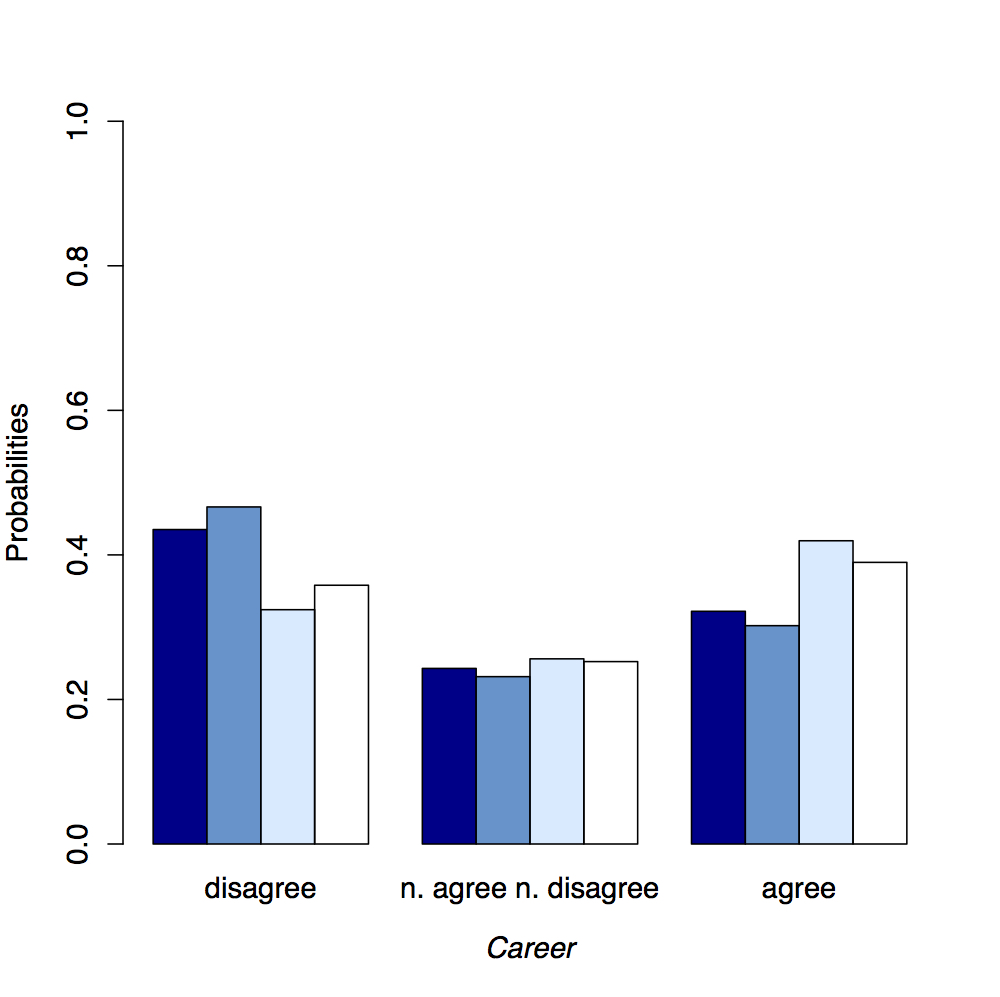}&
		\includegraphics[trim=0cm 1cm 1cm .5cm, width=6cm]{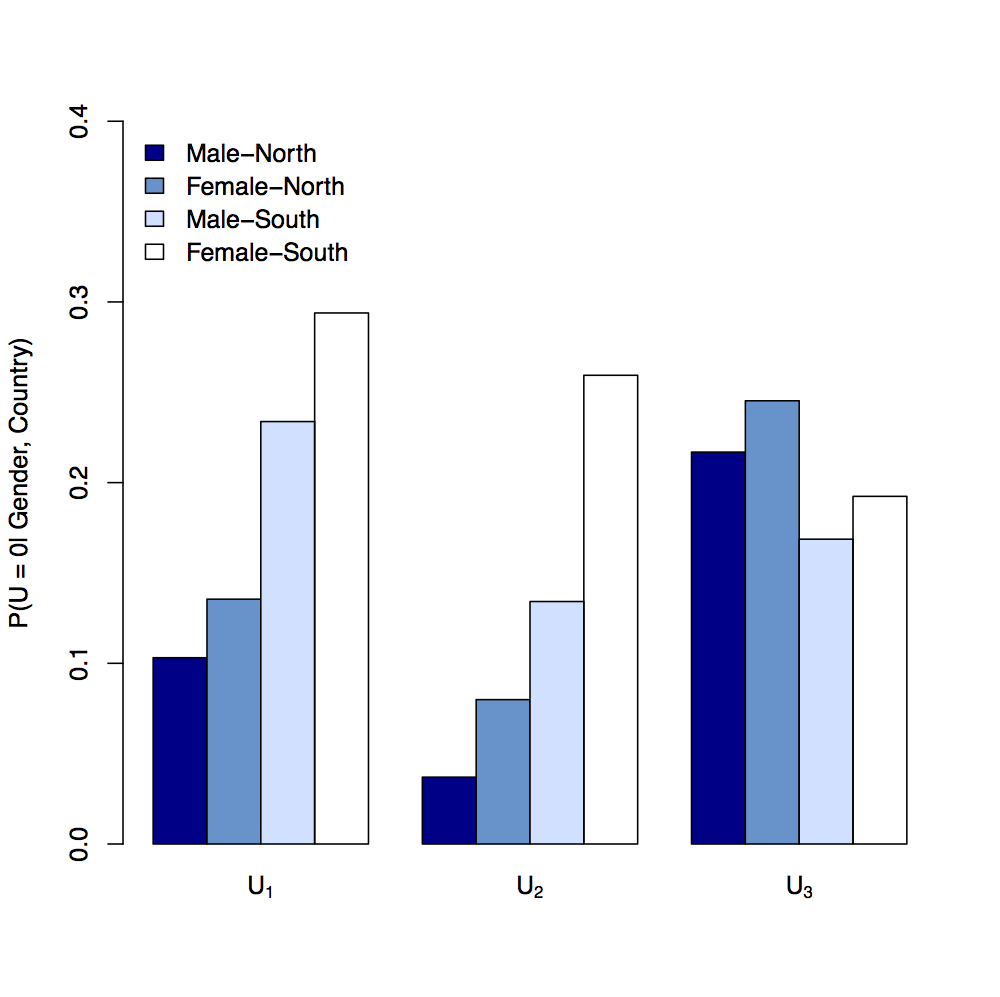}
	\end{tabular}
	\captionsetup{font=footnotesize,width=0.8\textwidth}\caption{Fitted distributions of the aware responses to \emph{Losejob} (top-left),  \emph{Wellpaid} (top-right),  \emph{Career} (bottom-left) and probabilities of being uncertain ($U_l = 0$, $l = 1, 2, 3$) in answering the questions (bottom-right) according to the covariate strata (\emph{Male-North, Female-North, Male-South, Female-South}) }\label{probobs}
	
\end{figure}

The estimated parameters of model $\mathcal{M}_{5}$ are  reported in Tables \ref{stimeobs} and \ref{stimelat}.
In particular in Table \ref{stimeobs}, the estimates of the parameters in equation \eqref{obslogit} highlight how workers' perceptions of the three aspects of their job vary according to \emph{Gender} and \emph{Country}.  \emph{Gender} has the same impact on the responses, whereas logits decrease for Southern workers when the question is \emph{Losejob} and increase for \emph{Wellpaid} and \emph{Career}.
The corresponding fitted distributions are illustrated in Figure \ref{probobs} (top and bottom-left).

\begin{table}[h!]
\centering
\captionsetup{font=footnotesize,width=0.9\textwidth}
\caption{
 Maximum likelihood estimates (MLE)  of the parameters of the linear models \eqref{obslogit} under $\mathcal{M}_5$,  with their standard errors (s.e.) and $p$-values
} \label{stimeobs}
	\renewcommand{\arraystretch}{1.5}
	\setlength{\tabcolsep}{2.4 mm}
	\resizebox*{0.9\textwidth}{!}{\begin{tabular}{lccccccccccc}
			\hline
			&  &     \multicolumn{3}{c}{\emph{Losejob}}      &            \multicolumn{3}{c}{\emph{Wellpaid}}    &              \multicolumn{3}{c}{\emph{Career}}      &  \\ \hline\hline
			Parameters                   &  &   MLE   &               s.e.                &  $p$-value       &   MLE   &              s.e.               &  $p$-value       &   MLE   &               s.e.               & $p$-value \\ \hline
			$\beta_i(1)$      &  & -1.5571
			 &            0.0655 &	0.0000
			       &0.2503 &	0.0619 &	0.0001 &
			            -0.3579 &	0.1470 &	0.0149
			\\
			$\beta_i(2)$      &  & -0.2918 &	0.1440 &	0.0427 &	0.1569 &	0.0476 &	0.0010 &	-0.0586 &	0.1462 &	0.6885
			\\
			$\beta_{i1}^{G}$                &  & -0.2243 &	0.0455 &	0.0000 &	-0.2252 & 0.0271 &	0.0000 &	-0.1088 &	0.0267 &	 0.0000
			  \\
			$\beta_{i1}^{C}$                &  & -0.1893  &	0.0646 &	0.0034 &	0.6288 &	0.0371 &	0.0000 &	0.3732 &	0.0327 &	 0.0000
			         \\ \hline
			&  & \multicolumn{3}{c}{\emph{Losejob,Wellpaid}} &          \multicolumn{3}{c}{\emph{Losejob,Career}} &          \multicolumn{3}{c}{\emph{Wellpaid,Career}} &  \\ \hline
			$\eta^{\{R_i, R_j\mid \b u^*\}}_{(1,1)}$ &  & 0.0826  &	0.0898 &	0.3572 &	0.3370 &	0.1034 &	0.0011 &	1.3550 &	0.1444 &	 0.0000 &\\
			$\eta^{\{R_i, R_j\mid \b u^*\}}_{(1,2)}$&  &-1.2757 &	0.2092 &	0.0000 &	-1.4139 &	0.2787 &	0.0000 &	0.3530  &	0.1180 &	 0.0028 \\$\eta^{\{R_i, R_j\mid \b u^*\}}_{(2,1)}$ &&
			-0.8518 &	0.1046 &	0.0000 &	-0.6649 &	0.1207 &	0.0000 &	7.0451 &	2E+02 &	0.9799 \\ $\eta^{\{R_i, R_j\mid \b u^*\}}_{(2,2)}$ &&
			-0.5245 &	0.4309 &	0.2234	 &-8.1503 &	1E+03 &	0.9960 &	1.5291 &	0.1256 &	0.0000	
			  \\ \hline
	\end{tabular}}
\vskip.5cm
\end{table}

\begin{table}[h!]
	\centering
	\captionsetup{font=footnotesize,width=0.9\textwidth}
	\caption{Maximum likelihood estimates (MLE)  of the parameters of the linear models \eqref{latlogit} under $\mathcal{M}_5$,  with their standard errors (s.e.) and $p$-values} \label{stimelat}
	\renewcommand{\arraystretch}{1.5}
	\setlength{\tabcolsep}{2.4 mm}
	\resizebox*{0.9\textwidth}{!}{\begin{tabular}{lccccccccccc}
			\hline
			&    &   \multicolumn{3}{c}{$U_1$}  &                \multicolumn{3}{c}{$U_2$}             &    \multicolumn{3}{c}{$U_3$}      &  \\ \hline\hline
			Parameters                   &  &   MLE   &               s.e.                &  $p$-value       &   MLE   &              s.e.               &    $p$-value     &   MLE   &               s.e.               & $p$-value \\ \hline
			$\tilde{\beta}_i$      &  & 2.1636 &	0.4410&	0.0000&	3.2589&	0.5784&	0.0000&	1.2839&	0.2995&	0.0000
			 \\
			$\tilde{\beta}_{i1}^{G}$                &  & -0.3103	&0.1847	&0.0930&	-0.8157&	0.1575&	0.0000&	-0.1601&	0.1236&	0.1953
			\\
			$\tilde{\beta}^{C}_{i1}$                &  & -0.9766 &	0.2147&	0.0000&	-1.3943&	0.4931&	0.0047&	0.3109&	0.2113&	0.1411
			    \\ \hline
	\end{tabular}}
\end{table}

The estimated log odds ratios in the last rows of Table \ref{stimeobs} suggest that aware responses on \emph{Wellpaid} and \emph{Career} are quite positively associated. On the contrary, as reasonably expected, \emph{Losejob}, is mainly negatively associated with the other responses. This result seems reasonable since workers who are worried about the loss of their job, probably do not meet good opportunities in their career or satisfaction in their remuneration.

The  estimates  of the parameters in equation \eqref{latlogit},  reported in Table \ref{stimelat}, and the corresponding probabilities in Figure~\ref{probobs} (bottom-right) show how the propensity to giving uncertain responses on \emph{Losejob} and \emph{Wellpaid} differs between male and female and among countries.
The question on \emph{Career} advancements has high proportion of uncertain responses, but women living in the South of EU tend to give uncertain answer more than the others when evaluating how plausible is losing their job.

The estimated shape parameters of the Local Reshaped Parabolic distributions that  model $R_i$ conditionally on  $U_i=0$, $i =1, 2, 3$ are all negative ($\widehat{\phi}_1 = -3.522$, $\widehat{\phi}_2 = -7.813$, $\widehat{\phi}_3 = -7.846$),  corresponding to  U-shaped distributions. This suggests that people giving uncertain answers tend to split into optimistic and pessimistic behaviours.  As for the comparison between model $\mathcal{M}_5$ and the analogous  estimated under the constraints $\widehat{\phi}_i=0, \: i=1,2,3,$  the test results $LRT=24.93$, the hypothesis of Uniform distribution for the uncertain responses has to be rejected.

\newpage

\section{Concluding remarks }\label{Conclusion}

The proposed  mixture model, HMMLU, is able to distinguish two kinds (awareness and uncertainty) of behaviour
that people may adopt, even unconsciously, when faced with  rating questions. It  allows to study the distribution of the aware responses and their dependence on covariate and  to model the association among responses given without uncertainty. Moreover, the HMMLU enables to specify different association structures among the binary latent variables governing the aware/uncertain behaviours and their dependence on covariates. As shown in Section~\ref{MonteCarlo}, ignoring uncertainty can result in erroneous estimates both of the rating distribution and the association parameters.

To model the uncertain responses, we introduce  a class of  distributions with a shape parameter that models different response styles and admits the Uniform distribution as a special case.
Nonetheless, the problem of selecting an adequate distribution of uncertainty is still an open problem and deserves further research.

A second critical aspect is that general results on  identifiability are lacking for HMMLU as for many other latent variable models. The issue is discussed in Sections~\ref{Sec.parameterization}, \ref{Sec.ident} and Appendix A, where we provided some necessary conditions. 
Empirical evidence on local identifiability, at least in a neighbourhood of the maximum likelihood estimate, is based on the fact that the Fisher matrix was never singular in the numerical examples and simulations we performed. Moreover,  a  data independent  assessment of local identifiability is provided by the numerical algorithm  described by \cite{forcina2008}.

Another point not considered in this paper is to test if uncertainty/awareness rules only some or none of the responses. Testing such hypotheses represents a non-standard problem as, under the null hypothesis, some parameters are on the boundary of their parametric space. For this reason, a comparison among HMMLU and models which do not contemplate uncertainty is not immediate. This problem can be solved, when the uncertain distribution is supposed to be Uniform, along the lines of \citet{colombigiordano} who dealt with the problem of testing uncertainty in a different multivariate model. However, the presence of shape parameters in the uncertainty distributions make the issue more complicated, since testing a no uncertainty hypothesis produces non-identifiability of these parameters. Such an issue deserves an in-depth study.

\vskip1.cm
\noindent \textbf{\large Acknowledgments}\\ The authors are grateful to Alan Agresti  for his thoughtful and suggestive comments.


\section*{Appendix A: Inference on marginal parameters}\label{Sec_MLest}

Analytical details to make inference on the marginal parameters of HMMLU models are here provided. 

Let $\b p_h$ be the vector of the joint probabilities  of the  configurations $\b r$ of the observable variables and the configurations $\b u$ of the latent ones in the $h^{th}$ covariate stratum, $h = 1, 2,\dots, H$. 

A marginal  parameterization of $\b p_h$ in terms of a vector of generalized marginal interactions $\b \eta_h$ is defined by the one-to-one mapping  $\b \eta_h=\b C \ln \b M \b p_h$ \citep{lang1994}. Here $\b C$ is a matrix of row contrasts and $\b M$ a matrix of $0$ and $1$ values to determine the marginal probabilities of interest \citep{bartolucci2007}.  Specifically, the marginal interactions in $\b \eta_h$ are contrasts of logarithms of sums of probabilities in $\b p_h$ (logits, log odds ratios, of any type, and contrasts of them).

Calculations are mainly based on the key result by 
\cite{bartolucci2007} that the transformation $\b \eta_h=\b C \ln \b M \b p_h$ is a diffeomorphism from the parameters $\b \theta_h$ of the saturated log-linear model for $\b p_h$ 
$$\b p_h=\frac{\exp(\b Z \b \theta_h)}{\b 1\tr  \exp(\b Z \b \theta_h)},$$
to the interactions $\b \eta_h$. Here $\b Z$ is the design matrix. 

For every  non-empty subset $\cg I$ of $\cg R \cup \, \cg U$, let  $\b \eta_h^{\cgl I}$ be the sub-vector of $\b \eta_h$ of the generalized interactions involving only variables in the set $\cg I$.
In the proposed  parameterization, every   $\b \eta_h^{\cgl I}$, $\cg I \subseteq \cg U$, is a vector of interactions  defined in the marginal distribution of the variables belonging to $\cg I$. These are   Glonek-McCullagh interactions that parameterize the vector  $\b \pi_h=(\b I_{2^v}\otimes \b 1_{m}\tr)\b p_h$ of the  joint probabilities of the latent variables.
Moreover, to  assure the smoothness of the parameterization \citep{bergsma2002}, every vector of interactions   $\b \eta_h^{\cgl I}$,
$\cg I \in \{\cg R_{\cgl S} \cup\,  \cg U_{\cgl T} :  \emptyset \subset \cgl S \subseteq \cgl V, \cgl T \subseteq \cgl V \}$, is defined in the marginal distribution of the variables in the set  $(\cg I \cap \cg R) \cup \: \cg U$.
These interactions parameterize the vector $\Diag(\b \pi_h\otimes \b 1_{m})^{-1}\b p_h$ of the probabilities of the responses given the latent variables.
All the previous interactions are defined by taking $0$ as the reference or base-line category of the logits of the latent variables (see \cite {colombi2014hmmm} for a description of how interactions are built starting from the logit types assigned to the variables).

Assumptions A1-A3 make some interactions  $\b \eta_h^{\cgl I}$ null. Interactions
defined in  the joint distribution of the variables in the sets  $ \{R_i\} \cup \: \cg U$ and $\{R_i,R_j\} \cup \; \cg U$, $i \neq j$, $i,j=1,2,\dots,v$, and those $\b \eta_h^{\cgl I}$, $\cg I \in \{\cg U_{\cgl T}:  \cgl T \subseteq \cgl V\}$, defined in the marginal distribution of the variables belonging to $\cg I$, are the only ones not constrained to be zero and correspond to the parameters involved in $i)-iv)$ of Theorem \ref{teo4}.

We can express therefore the parameter constraints through the linear model $\b \eta_h=\b X_h \b \beta$, $h = 1, 2,\dots, H$, where $\b \beta$ is the vector of unknown parameters, including the shape parameters of the Reshaped Parabolic distributions. This linear model  accounts for the dependence structure of both latent and observed variables and the effects of covariates. 

We now provide the analytical details for the ML estimation of 
$\b \beta$. To this regard, we utilize some results   by \citet{forcina2008}, as  the HMMLU  of Section \ref{Sec.hierarch} can be viewed as a special case of  his \emph{Extended Latent Class Model}. 

We start from the mentioned diffeomorphism $\b \eta_h=\b C \ln \b M \b p_h$ to  obtain 
 $$\b R_h = \frac{\partial \b \theta_h}{\partial \b \eta_h \tr}=\left ( \b C \; {\Diag}^{-1}(\b M \b p_h) \; \b M \, \b \Omega_h \b Z \right)^{-1},$$
with $\b \Omega_h= {\Diag}(\b p_h)- \b p_h \b p_h\tr$.

Moreover,  denote the saturated log-linear model for the vector $\b q_h$ of the joint probabilities of the responses in the $h^{th}$ stratum by
$$\b q_h= \b L \b p_h=\frac{\exp(\b W \b \gamma_h)}{\b 1\tr  \exp(\b W \b \gamma_h)},$$
where $\b L$ is the marginalization matrix with respect to the latent variables, $\b W$ is the design matrix of the log-linear model and $\b \gamma_h=\b H \ln\left(\b L \b p_h\right)$ is a vector of contrasts of logarithms of the elements of $\b q_h$, with $\b H \b W = \b I_{m-1}$.
By the chain rule of matrix differential calculus \citep{magnus2007}, we get
$\b D_h=\frac{\partial \b \gamma_h}{\partial \b \beta \tr}=\b Q_h \b R_h \b X_h,$
where $\b Q_h=\frac{\partial \b \gamma_h}{\partial \b \theta_h \tr}=\b H {\Diag}^{-1}(\b q_h) \; \b L \, \b \Omega_h \b Z.$
It also follows that
$\frac{\partial \b q_h}{\partial \b \beta\tr}=\left ({\Diag}(\b q_h)- \b q_h \b q_h\tr \right) \, \b W \, \b D_h,$
which is the main result needed for calculating the Fisher matrix.

Let  $\b n_h$  indicate the observed joint frequencies of the responses in the $h^{th}$ stratum of size $n_h$ and  $n=\sum_{h=1}^H n_h$ be the total sample size.
Under multinomial sampling within every stratum, the log-likelihood function is
$L_n=\sum_{h=1}^H \b n_h \tr \ln (\b q_h),$
and the row vector of the score functions is
$ \b S_n=\sum_{h=1}^H (\b n_h-n_h \b q_h)\tr \, \b W \, \b D_h.$
From the previous results, the averaged Fisher matrix  easily follows $\b F_n= \frac{1}{n}  E(\b S_n \tr \b S_n)=
\frac{1}{n} \displaystyle \sum_{h=1}^H \b D_h\tr \b W \tr \left (n_h  {\Diag}(\b q_h)- n_h\b q_h \b q_h\tr \right)\b W \b D_h.$
If  $\displaystyle \lim_{n\rightarrow\infty}\frac{n_h}{n} = \omega_h>0, \:h=1,2,\dots,H,$ then
$\b F= \lim_{n\rightarrow\infty}\b F_n= \sum_{h=1}^H \omega_h \b D_h\tr \b W \tr \left (\Diag(\b q_h)- \b q_h \b q_h\tr \right )\b W \b D_h.$

Since $\b W \tr (\omega_h  \Diag(\b q_h)- \omega_h\b q_h \b q_h\tr)\b W $ is non singular,  $\b F$ is  non singular if and only if
the $H(m-1)\times p$ matrix $\b D$, obtained by row-binding the matrices $\b D_h$, is of full column rank.  Thus  $rank(\b D)=p$ implies that  the vector of parameters $\b \beta$, at which $\b D$ is computed, is locally identifiable \citep{Roth, forcina2008}.

Hence, $\b \beta$ denotes the vector of the true parameters and $\b p, \b q, \b \pi$, $\b D$ and the other just introduced matrices, will be computed at this value.
If $rank(\b D)=p$, from the standard MLE theory, it follows that $\widehat{\b\beta} -\b \beta $ has an asymptotic Normal distribution with null expected value and variance matrix $\frac{1}{n} \b F^{-1}$.

\section*{Appendix B: Reshaped Parabolic distributions}
\begin{figure}[h!]
	\centering
	\begin{tabular}{p{7.5cm} | p{7.5cm}}
	\includegraphics[width=7.5cm]{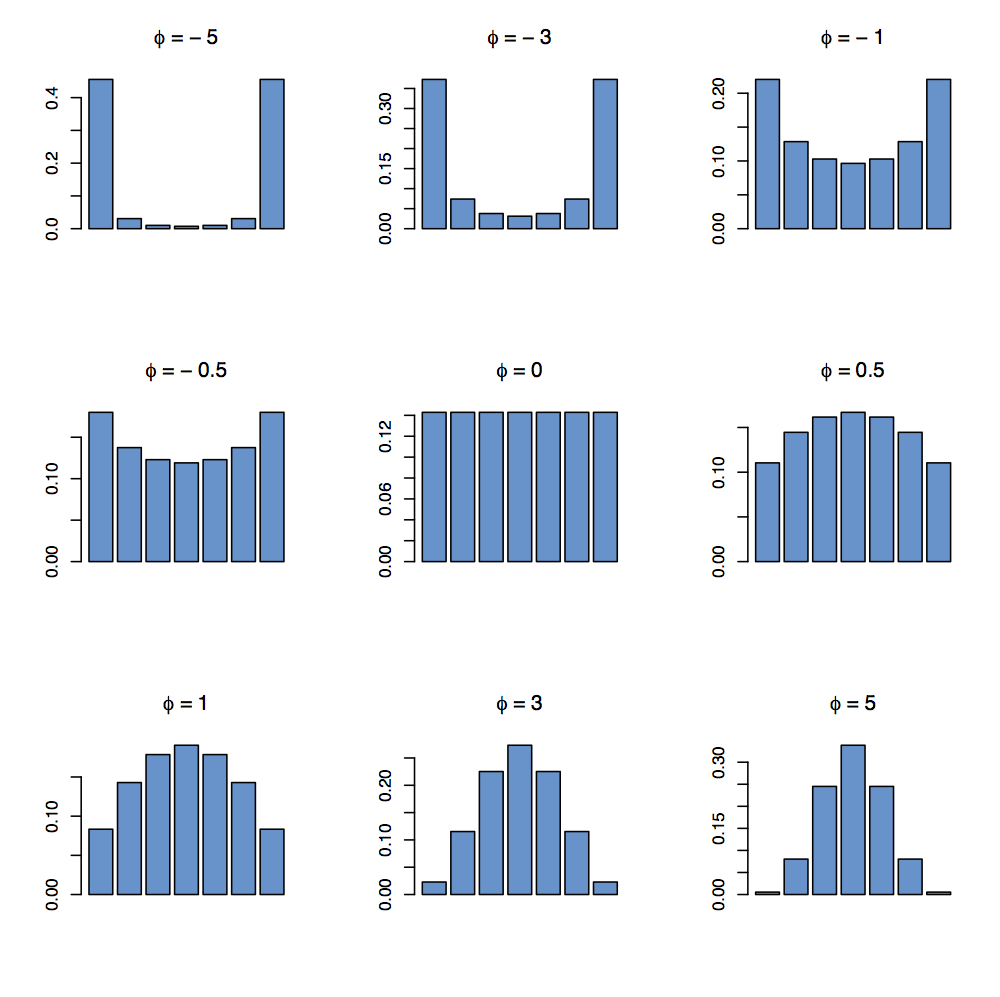}&
 \includegraphics[width=7cm]{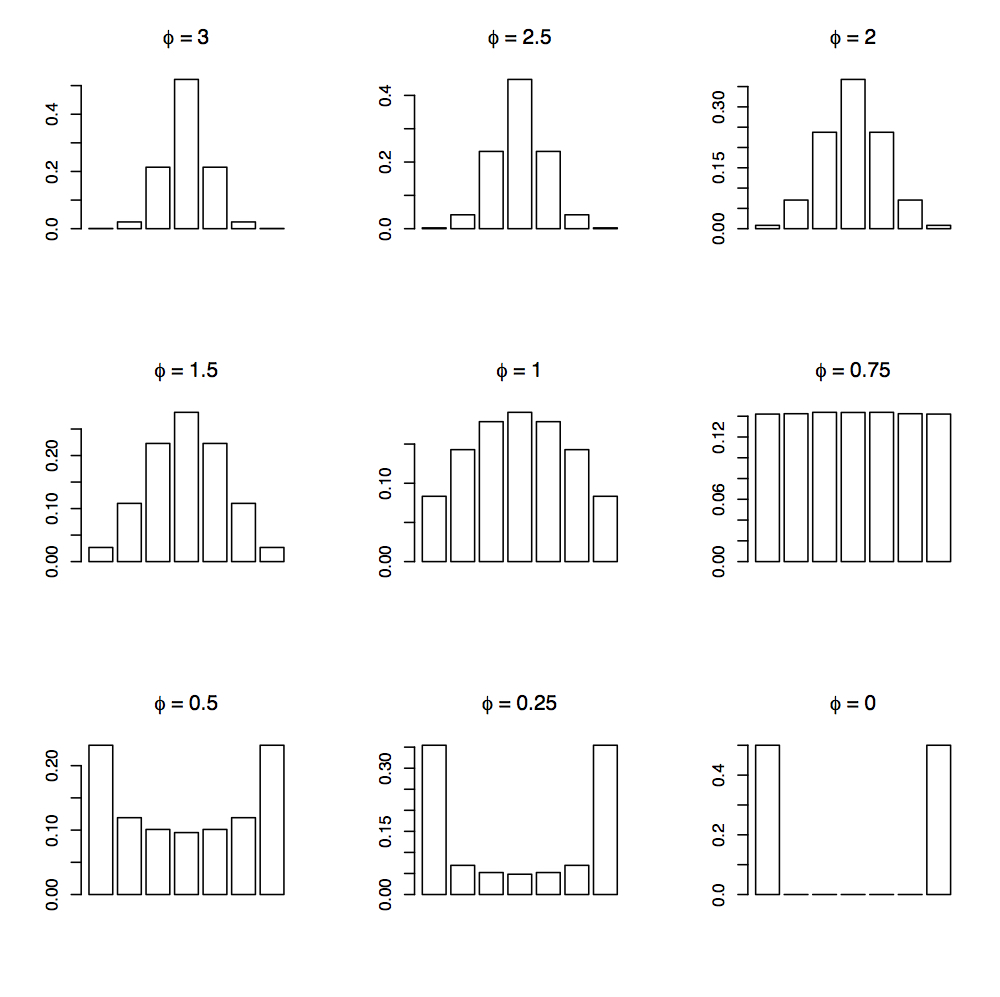}
\end{tabular}
\captionsetup{font=footnotesize,width=0.9\textwidth}\caption{Local (coloured) and Global (white) Reshaped Parabolic distributions with different shape parameters}\label{LRPD}
\end{figure}

Given an ordinal  categorical variable with $m$ categories, the Reshaped Parabolic distribution is defined by the  powers  $\left (\frac{p(r+1)}{p(r)}\right)^{\phi}$ of the local  odds or by the powers $\left(\frac{1-F(r)}{F(r)}\right)^{\phi}$  of the global odds, $r=1,2,...,m-1$, of the discrete Parabolic probability function
 $$p(r)=\frac{6(m+1-r)r}{(m+2)(m+1)m}, \qquad r=1,2,\dots,m$$ with
 distribution function
 $$F(r)= \frac{r(r+1)(3(m+1)-2r-1)}{(m+2)(m+1)m}, \qquad r=1,2,\dots,m.$$
Local and global odds lead to two different Reshaped Parabolic probability functions  which will be called Local and Global {Reshaped}, respectively.
The Local {Reshaped} Parabolic distribution family contains, as a special case with $\phi=0$, the Uniform distribution, for positive  $\phi$ it is bell shaped  and for negative $\phi$ it is U-shaped.
The Global Reshaped Parabolic distribution is defined only for $\phi \geq 0$ and assigns probability $1/2$ to the two extreme categories when $\phi=0$. For  $\phi\geq 1$ it is bell shaped  and U-shaped for  $\phi<1 $.
Both Local and Global Reshaped Parabolic are symmetric, have expected value independent of $\phi$ and variance which is a decreasing function of $\phi$.
Figure~\ref{LRPD} shows some examples.

 Distributions similar to the Reshaped Parabolic can be obtained
from the powers of logits of other symmetric probability functions which do not
depend on unknown parameters such as, for example, the Triangular probability function
$$p(r)=\frac{2}{m+1}\left \{\frac{r} {  \frac {m+1}{ 2 }   }\, \delta_{(r<  \frac {m+1}{ 2 } )}+\frac{m-r+1}{m-  \frac {m+1}{ 2 } +1}\, \delta_{(r \geq \frac {m+1}{ 2 })  }\right \},\: \: \: r=1,2,\dots,m$$
where $\delta_{(.)}$ is the Dirac measure.
\bibliographystyle{chicago}
\bibliography{mybibnew}

\begin{thebibliography}{}

\bibitem[\protect\citeauthoryear{Bartolucci, Colombi, and Forcina}{Bartolucci
  et~al.}{2007}]{bartolucci2007}
Bartolucci, F., R.~Colombi, and A.~Forcina (2007).
\newblock An extended class of marginal link functions for modelling
  contingency tables by equality and inequality constraints.
\newblock {\em Statistica Sinica\/}~{\em 17}, 691--711.

\bibitem[\protect\citeauthoryear{Baumgartner and Steenkamp}{Baumgartner and
  Steenkamp}{2001}]{baumgartner2001}
Baumgartner, H. and J.~B.~E. Steenkamp (2001).
\newblock Response styles in marketing research: a cross-national
  investigation.
\newblock {\em Journal of Marketing Research\/}~{\em 38\/}(2), 143--156.

\bibitem[\protect\citeauthoryear{Bergsma and Rudas}{Bergsma and
  Rudas}{2002}]{bergsma2002}
Bergsma, W.~P. and T.~Rudas (2002).
\newblock Marginal models for categorical data.
\newblock {\em Annals of Statistics\/}~{\em 30\/}(1), 140--159.

\bibitem[\protect\citeauthoryear{B{\"o}ckenholt and Meiser}{B{\"o}ckenholt and
  Meiser}{2017}]{bockenholt2017}
B{\"o}ckenholt, U. and T.~Meiser (2017).
\newblock Response style analysis with threshold and multi-process {IRT}
  models: a review and tutorial.
\newblock {\em British Journal of Mathematical and Statistical
  Psychology\/}~{\em 70\/}(1), 159--181.

\bibitem[\protect\citeauthoryear{Colombi and Giordano}{Colombi and
  Giordano}{2016}]{colombigiordano}
Colombi, R. and S.~Giordano (2016).
\newblock A class of mixture models for multidimensional ordinal data.
\newblock {\em Statistical Modelling\/}~{\em 16\/}(4), 322--340.

\bibitem[\protect\citeauthoryear{Colombi, Giordano, and Cazzaro}{Colombi
  et~al.}{2014}]{colombi2014hmmm}
Colombi, R., S.~Giordano, and M.~Cazzaro (2014).
\newblock hmmm: an {R} package for hierarchical multinomial marginal models.
\newblock {\em Journal of Statistical Software\/}~{\em 59\/}(11), 1--25.

\bibitem[\protect\citeauthoryear{D'Elia and Piccolo}{D'Elia and
  Piccolo}{2005}]{delia2005}
D'Elia, A. and D.~Piccolo (2005).
\newblock A mixture model for preferences data analysis.
\newblock {\em Computational Statistics \& Data Analysis\/}~{\em 49\/}(3),
  917--934.

\bibitem[\protect\citeauthoryear{Forcina}{Forcina}{2008}]{forcina2008}
Forcina, A. (2008).
\newblock Identifiability of extended latent class models with individual
  covariates.
\newblock {\em Computational Statistics \& Data Analysis\/}~{\em 52\/}(12),
  5263--5268.

\bibitem[\protect\citeauthoryear{Fr\"{u}wirth~Schnatter}{Fr\"{u}wirth~Schnatter}{2006}]{Silvia2006}
Fr\"{u}wirth~Schnatter, S. (2006).
\newblock {\em Finite Mixture and Markov Switching Models}.
\newblock Springer, New York, USA.

\bibitem[\protect\citeauthoryear{Glonek and McCullagh}{Glonek and
  McCullagh}{1995}]{glonek1995}
Glonek, G.~F. and P.~McCullagh (1995).
\newblock Multivariate logistic models.
\newblock {\em Journal of the Royal Statistical Society. Series B
  (Methodological)\/}~{\em 57\/}(3), 533--546.

\bibitem[\protect\citeauthoryear{Gottard, Iannario, and Piccolo}{Gottard
  et~al.}{2016}]{gottard2016}
Gottard, A., M.~Iannario, and D.~Piccolo (2016).
\newblock Varying uncertainty in {CUB} models.
\newblock {\em Advances in Data Analysis and Classification\/}~{\em 10\/}(2),
  225--244.

\bibitem[\protect\citeauthoryear{Hojsgaard}{Hojsgaard}{2004}]{H2004}
Hojsgaard, S. (2004).
\newblock Statistical inference in context specific interaction models for
  contingency tables.
\newblock {\em Scandinavian Journal of Statistics\/}~{\em 31\/}(1), 143--158.

\bibitem[\protect\citeauthoryear{Huang}{Huang}{2016}]{huang2016}
Huang, H.~Y. (2016).
\newblock Mixture random-effect {IRT} models for controlling extreme response
  style on rating scales.
\newblock {\em Frontiers in Psychology\/}~{\em 7}, 1706.

\bibitem[\protect\citeauthoryear{Jin and Wang}{Jin and Wang}{2014}]{jin2014}
Jin, K.~Y. and W.~C. Wang (2014).
\newblock Generalized {IRT} models for extreme response style.
\newblock {\em Educational and Psychological Measurement\/}~{\em 74\/}(1),
  116--138.

\bibitem[\protect\citeauthoryear{Kateri}{Kateri}{2014}]{kateri2014book}
Kateri, M. (2014).
\newblock {\em Contingency Table Analysis: {Methods} and Implementation Using
  {R}}.
\newblock Springer.

\bibitem[\protect\citeauthoryear{Kauermann}{Kauermann}{1997}]{kauermann1997}
Kauermann, G. (1997).
\newblock A note on multivariate logistic models for contingency tables.
\newblock {\em Australian \& New Zealand Journal of Statistics\/}~{\em
  39\/}(3), 261--276.

\bibitem[\protect\citeauthoryear{Lang and Agresti}{Lang and
  Agresti}{1994}]{lang1994}
Lang, J.~B. and A.~Agresti (1994).
\newblock Simultaneously modeling joint and marginal distributions of
  multivariate categorical responses.
\newblock {\em Journal of the American Statistical Association\/}~{\em
  89\/}(426), 625--632.

\bibitem[\protect\citeauthoryear{Luchini and Watson}{Luchini and
  Watson}{2013}]{luchini2013}
Luchini, S. and V.~Watson (2013).
\newblock Uncertainty and framing in a valuation task.
\newblock {\em Journal of Economic Psychology\/}~{\em 39}, 204--214.

\bibitem[\protect\citeauthoryear{Lupparelli, Marchetti, and Bergsma}{Lupparelli
  et~al.}{2009}]{lupparelli2009}
Lupparelli, M., G.~M. Marchetti, and W.~P. Bergsma (2009).
\newblock Parameterization and fitting of bi-directed graph models to
  categorical data.
\newblock {\em Scandinavian Journal of Statistics\/}~{\em 36\/}(3), 559--576.

\bibitem[\protect\citeauthoryear{Magnus and Neudecker}{Magnus and
  Neudecker}{2007}]{magnus2007}
Magnus, J.~R. and H.~Neudecker (2007).
\newblock {\em Matrix Differential Calculus with Applications in Statistics and
  Econometrics. {Third} edition.}
\newblock John Wiley \& Sons.

\bibitem[\protect\citeauthoryear{Meade and Craig}{Meade and
  Craig}{2012}]{meade2012}
Meade, A.~W. and S.~B. Craig (2012).
\newblock Identifying careless responses in survey data.
\newblock {\em Psychological methods\/}~{\em 17\/}(3), 437--455.

\bibitem[\protect\citeauthoryear{Morren, Gelissen, and Vermunt}{Morren
  et~al.}{2011}]{morren2011}
Morren, M., J.~P. Gelissen, and J.~K. Vermunt (2011).
\newblock Dealing with extreme response style in cross-cultural research: a
  restricted latent class factor analysis approach.
\newblock {\em Sociological Methodology\/}~{\em 41\/}(1), 13--47.

\bibitem[\protect\citeauthoryear{Rothenberg}{Rothenberg}{1971}]{Roth}
Rothenberg, T. (1971).
\newblock Identification in parametric models.
\newblock {\em Econometrica\/}~{\em 39}, 577--591.

\bibitem[\protect\citeauthoryear{Studeny}{Studeny}{2005}]{studeny2005}
Studeny, M. (2005).
\newblock {\em Probabilistic Conditional Independence Structures}.
\newblock London, Springer.

\bibitem[\protect\citeauthoryear{Tutz, Schauberger, and Berger}{Tutz
  et~al.}{2018}]{tutz2018}
Tutz, G., G.~Schauberger, and M.~Berger (2018).
\newblock Response styles in the partial credit model.
\newblock {\em Applied Psychological Measurement\/}, OnlineFirst,
  \emph{doi.org/10.1177/0146621617748322}.

\bibitem[\protect\citeauthoryear{von Davier and Yamamoto}{von Davier and
  Yamamoto}{2007}]{von2007mixture}
von Davier, M. and K.~Yamamoto (2007).
\newblock Mixture-distribution and hybrid rasch models.
\newblock In M.~von Davier and C.~H. Carstensen (Eds.), {\em Multivariate and
  Mixture Distribution Rasch Models}, pp.\  99--115. Springer.

\bibitem[\protect\citeauthoryear{Yates, Lee, and Bush}{Yates
  et~al.}{1997}]{yates1997}
Yates, J.~F., J.~W. Lee, and J.~G. Bush (1997).
\newblock General knowledge overconfidence: cross-national variations, response
  style, and ``reality".
\newblock {\em Organizational Behavior and Human Decision Processes\/}~{\em
  70\/}(2), 87--94.

\end{thebibliography}

\end{document}